\documentclass[lettersize,journal]{IEEEtran}
\usepackage{amsmath,amsfonts}
\usepackage{algorithmic}
\usepackage[linesnumbered, ruled]{algorithm2e}
\usepackage{array}
\usepackage[caption=false,font=normalsize,labelfont=sf,textfont=sf]{subfig}
\usepackage{textcomp}
\usepackage{stfloats}
\usepackage{url}
\usepackage{verbatim}
\usepackage{graphicx}
\usepackage{cite}
\usepackage{dsfont}
\usepackage{multirow}
\usepackage{makecell}
\usepackage{color,soul}
\usepackage{lineno}

\begin{document}

\title{Bayesian Non-parametric Hidden Markov Model for Agile Radar Pulse Sequences Streaming Analysis}

\author{Jiadi Bao,~\IEEEmembership{Student Member,~IEEE,}
Yunjie Li,~\IEEEmembership{Senior Member,~IEEE,}\\
Mengtao Zhu,~\IEEEmembership{Member,~IEEE,}
and Shafei Wang
\thanks{This work was supported by the National Natural Science Foundation (NSFC) of China under grants no. 61976019. \textit{(corresponding author: Mengtao Zhu)}.}
\thanks{Jiadi Bao is with the School of Information and Electronics, Beijing Institute of Technology, Beijing, 100081, China (E-mail: baojiadi@bit.edu.cn).}
\thanks{Yunjie Li is with the School of Information and Electronics, Beijing Institute of Technology, Beijing, 100081, China, and also with the Peng Cheng Laboratory, Shenzhen, 518055, China (E-mail: liyunjie@bit.edu.cn).}
\thanks{Mengtao Zhu is with the School of Cyberspace Science and Technology, Beijing Institute of Technology, Beijing, 100081, China, and also with the Peng Cheng Laboratory, Shenzhen, 518055, China (E-mail: zhumengtao@bit.edu.cn). }
\thanks{Shafei Wang is with the School of Information and Electronics, Beijing Institute of Technology, Beijing, 100081, China, and also with the Laboratory of Electromagnetic Space Cognition and Intelligent Control, Beijing, 100191, China (E-mail: rockingsand-storm@163.com).}
}
\markboth{Journal of \LaTeX\ Class Files,~Vol.~14, No.~8, May~2023}%
{Shell \MakeLowercase{\textit{et al.}}: A Sample Article Using IEEEtran.cls for IEEE Journals}

\IEEEpubid{0000--0000/00\$00.00~\copyright~2021 IEEE}

\maketitle

\begin{abstract}
Multi-function radars are sophisticated types of sensors with the capabilities of complex agile inter-pulse modulation implementation and dynamic work mode scheduling. The developments in MFRs pose great challenges to modern electronic reconnaissance systems or radar warning receivers for recognition and inference of MFR work modes. To address this issue, this paper proposes an online processing framework for parameter estimation and change point detection of MFR work modes. At first, this paper designed a fully-conjugate Bayesian Non-Parametric Hidden Markov Model with a designed prior distribution (agile BNP-HMM) to represent the MFR pulse agility characteristics. Then, the proposed framework is constructed by two main parts. The first part is the agile BNP-HMM model for automatically inferring the number of HMM hidden states and emission distribution of the corresponding hidden states. An error lower bound is derived for estimation performance and the proposed algorithm is shown to be closer to the bound compared with baseline methods. The second part combines the streaming Bayesian updating to facilitate computation, and designed an online work mode change detection framework based upon the weighted sequential probability ratio test. We demonstrate that the proposed framework is consistently highly effective and robust to baseline methods on diverse simulated radar signal data and real-life benchmark datasets. 
\end{abstract}


\section{Introduction}
Multi-Function Radars (MFRs) can schedule multiple simultaneous work modes for different tasks in the radar timeline~\cite{wang_signal_2008, visnevski_syntactic_2007, visnevski_syntactic_2005}, such as surveillance, target tracking, and track maintenance. With increased agility, MFRs can adapt and optimize their control parameters (such as Pulse Repetition Interval (PRI), Radio Frequency (RF), and Pulse Width (PW)) based on some mission-specific objectives~\cite{haigh_cognitive_2021}. The increased flexibility and adaptability of MFRs pose great challenges to modern Electronic Support (ES) systems~\cite{wiley_elint_2006,wang_threat_2007,arasaratnam_tracking_2006}. To protect a target from the radar threat, the ES needs to recognize dynamic and complex work modes with possibly little prior information and detect the MFR work mode transition as soon as possible. Such that effective tactical and jamming countermeasures can be scheduled.

Traditionally, the recognition of MFR work mode is based on the supervised method. Various MFR models and recognition methods have been designed~\cite{apfeld_modelling_2019, li_work_2020,xueqiong_toward_2018,liu_classification_2019}. However, due to the agility and software defined ability of modern MFR \cite{huang_analysis_2018, li_phase_2021}, the prior information required for previous supervised methods for recognition is difficult to obtain or lose efficacy as new radars and radar signals always exists in the dynamic electromagnetic environments. To alleviate the requirements in prior information, model based time series clustering methods is proposed for unsupervised recognition and inter-pulse modulation parameter estimation of MFR pulse sequences \cite{zhu_model-based_2021}.  However, the methods in \cite{zhu_model-based_2021} required further algorithmic designs for non-ideal conditions caused by spurious pulses and missing pulses in real world electromagnetic environments. In addition, facing the work modes with dynamic and agile inter-pulse modulations, they introduced a search-based method to find the true number of PRI levels \cite{zhu_model-based_2021}. This search operation is computationally expensive and is not suitable for online paradigm. It is necessary to develop methods for robust recognition and parameter estimation of different radar work modes especially under non-ideal situations and detecting the change point among consecutive work models in online manner.

\IEEEpubidadjcol


Quickest change point detection\cite{tartakovsky2014sequential} is a viable approach to solve the above motioned MFR problems. The first step is to effectively model the radar pulse sequence of different work modes. In general, the radar pulse sequences can be modeled though a parameterized model as a combination of inter and inner pulse modulations on multiple radar control parameters (such as PRI, RF and PW). For instance, in~\cite{zhu_model-based_2021}, four parametric models were used to represent pulse sequence with different inter-pulse modulations. In~\cite{scherreik_online_2021, revillon2019radar}, the Gaussian Mixture Model (GMM) was used to characterize the radar pulse sequences. In~\cite{wang_signal_2008,visnevski_syntactic_2005,visnevski_syntactic_2007}, hidden Markov model has been utilized for modeling pulse sequences of the well known ``Mercury'' multi-function radar\footnote{Different pulse sequences correspond to different ``radar words'' in their research.}. The HMM can map both the inter-pulse modulation type and modulation parameters of different radar pulse sequence to a unified state space model. Each hidden state in a HMM corresponds to a specific pulse parameter value and the influence of non-ideal conditions can be modeled through introducing external hidden states. One problem of HMM inference in non-cooperative environments is that the number of hidden states $K$ is unknown and needs to be specified. Besides, since the hidden states in HMM for radar pulse sequences have explicit physical meanings, improper settings of $K$ would definitely degrade the performance and reduce the interpretability~\cite{ding_variational_2010}. 


For a better representation model, the Bayesian Non-Parametric HMM (BNP-HMM) with infinite number of hidden state assumptions may serve as a potential candidate. In BNP-HMM each effective state corresponds to a hidden state with a relatively large posterior probability. The first attempts to use infinite number of hidden state assumption on HMM was performed in~\cite{teh_hierarchical_2006,beal_infinite_2001}. Later, a variety of investigations have been conducted based on infinite number of hidden state assumption, such as in speaker diarization~\cite{fox_sticky_2011}, radar High Resolution Range Profile (HRRP) recognition~\cite{du2011bayesian,chen2021tensor}, and human motion prediction~\cite{fox_joint_2014}. The Bayesian non-parametric theory provided the theoretical foundation in modeling the MFR pulse sequences and achieving parameter estimation with little prior information. To the best of the authors' knowledge, there has been no research of Bayesian non-parametric theory for intercepted radar signal analysis. Meanwhile, above mentioned investigations are based on non-agile or sticky assumptions about the hidden state transitions, and are unsuitable for estimating agile and dynamic MFR work-mode hidden state transitions with corruptions caused by non-ideal situations. Thus, for MFR applications, there are two problems that must be solved for existing BNP-HMM methods: 1) Specific designs are needed for the dynamic and agile property and non-ideal situations in MFR pulse sequences. 2) An online processing framework is needed that has efficient model learning and inference.

The second step for quickest change point detection of MFR work modes is to design the detection strategy. The quickest change point detection of the HMM model (actually the state space model) is challenging due to two main reasons: Firstly, if the pre-change distribution is known and the post-change distribution is unknown, estimating or even modeling the post-change distribution is often impractical as we may not know a priori what kind of change (including changes in structure and trend) will occur~\cite{tartakovsky_asymptotic_2019,fuh_asymptotic_2019, fuh_asymptotically_2021}. Secondly, computing the Kullback-Leibler (KL) divergence is necessary for evaluating the change point detection test. But calculating KL divergence in state space models is always non-trivial~\cite{fuh_sprt_2003, fuh_quickest_2015}. For HMM, a common method is to use Markov chain Monte Carlo simulations to approximate the KL divergence, and asymptotic properties can be derived\cite{fuh_asymptotic_2019}. Fuh \cite{fuh_sprt_2003} treated the HMM as a big Markov chain, and recursive CUmulative SUM (CUSUM) was formulated based on mini-max criterion. Fuh~\cite{fuh_quickest_2015} also derived the logarithmic likelihood and developed a non-Monte-Carlo method to compute KL divergence for two-state HMM. On the one hand, an effective numerical algorithm for KL divergence of any finite state HMM is remained to be solved as currently only two-state HMM was examined. On the other hand, the considered BNP-HMM for pre- and post-change distribution modeling is non-ergodic and has hidden state space with infinite cardinality, which makes the general results achieved in classical HMM not applicable.


Taking the above problems into consideration, this paper proposes a new framework that performs the Parameter Estimation task (PE task) and the Change Point Detection task (CPD task) of the MFR work modes. The framework is termed as Agile BNP-HMM-CUSUM and is consisted of three consecutive steps: MFR pulse sequence modeling, work mode parameter estimation, and change point detection. Firstly, for a better pulse sequence modeling a variant of the BNP-HMM, the agile BNP-HMM, is proposed to achieve improved control over the inferred HMM hidden states. Such control is crucial for the MFR problems we examine. Secondly, an efficient variational inference method is designed for the proposed agile BNP-HMM. The proposed method can automatically estimate the number of hidden states in BNP-HMM through the agile Dirichlet Process (DP). The agile DP is a measure on measures and is parameterized by a base distribution and a positive scaling parameter. To represent the designed agile DP, a corresponding agile stick-breaking construction method is proposed based on the conventional stick-breaking construction \cite{sethuraman1994constructive}. Our agile BNP-HMM model is further combined with streaming Bayesian updating~\cite{broderick_streaming_2013} processing. The streaming updating can mitigate the PE task from local optimal and facilitate computation. Finally, to overcome the computationally demanding CPD test of HMM due to the real-time processing requirements and maintain high performance, We consider the initial distributions and transition matrices as nuisance parameters, as done in \cite{fuh_quickest_2015}. Then the problem of detecting change points for HMM is transformed into the problem of detecting change points for multivariate Gaussian and a weighted Sequential Probability Ratio Test (SPRT) is proposed for MFR applications\footnote{This simplification allows us to make use of the results from previous research on detecting change points for multivariate Gaussians \cite{nikiforov_quadratic_1999}. Actually, similar simplifications can be found in Stephen Boyd's method \cite{hallac2017toeplitz} for Markov Random Field. In \cite{hallac2017toeplitz}, the highly structured variables are transformed into multivariate Gaussian inverse covariance matrices to simplify the raw problems with intensive computations.}. 

The main contributions of this work can be summarized as follows:
\begin{enumerate}
\item[1.] An agile BNP-HMM is proposed to model MFR work modes with agile and dynamic properties. The proposed model is fully conjugated and permits high efficient variational inference.
\item[2.] An improved  stick-breaking construction method is developed. Compared to the conventional stick-breaking construction method proposed by Sethuraman~\cite{sethuraman1994constructive}, our proposed method provides a more accurate estimation facing the agile inter-pulse modulation types. The error lower bound for estimation performance under non-ideal conditions is also derived.
\item[3.]An asymptotic optimal change point detection strategy is designed base upon a weighted Sequential Probability Ratio Test (weighted SPRT) \cite{basseville1993detection}. The designed strategy does not require pre-specified window size information required for sliding window based methods, and is thus free of the window adjustment dilemma and achieves a simpler parameter tuning result.
\item[4.] The proposed agile BNP-HMM-CUSUM is combined with streaming Bayesian update techniques \cite{broderick_streaming_2013} for online processing. The computational redundancy is reduced.
\end{enumerate}

The rest of the paper is organized as follows. Section~\ref{sec:problem formulation} describes the PE and CPD task formulations. Section~\ref{sec:method} introduces the proposed agile BNP-HMM-CUSUM framework for PE and CPD tasks. Section~\ref{sec:simulated performance on radar data} presents the simulation design, results, and discussions. Section~\ref{sec:experimental performance on real-life datasets} presents the experiments on real-life benchmark datasets. Section~\ref{sec:conclusion} concludes the paper. Main notations of this paper is shown in Table \ref{notations}.

\begin{table}
    \caption{Main notations used in this paper}
	\centering
    \label{notations}
	\begin{tabular}{ll}
	    \hline
	    \textbf{Notations} & \textbf{Descriptions}\\
	    \hline
        $\boldsymbol{P}^t$ & The data batch of index $t$ \\
        $\boldsymbol{S}$ & The underlying state sequence of HMM\\
        $\boldsymbol{\pi}$ and $\boldsymbol{A}$ & The initial and transition distribution of HMM\\
        $\boldsymbol{\Theta}_t$ & The set of Gaussian distributions at time $t$\\
        $\boldsymbol{\Upsilon}$ & The set of hyper-parameters of the BNP-HMM\\
        $\boldsymbol{Z}^t$ & The latent variable of data batch of index $t$\\
        $\boldsymbol{\varphi}^k$ & The parametric model of the \textit{k} th hidden state\\
        $T$ & The length of the data batch\\
        $L$ & The truncation level\\
        $K$ & The number of hidden states\\
        $N$ & The detected change point (stopping time)\\
        $\nu$ & The true change point\\
        $\boldsymbol{\theta}_0$ and $\boldsymbol{\theta}_1$ & The pre- and post-change distribution parameter\\
        $\mathbb{E}_q$ & The expectation with respect to distribution $q$ \\
        \hline
	\end{tabular}
\end{table}

\section{Problem Formulation}
\label{sec:problem formulation}
Without losing the generality, this paper defines the PE and CPD tasks of the MFR work modes by PRI parameters, but the proposed method can be extended to other Pulse Descriptive Word (PDW) and multi-parameters case. Firstly, a probabilistic generative model for different PRI modulations is presented. Then, the mathematical formulations of the PE and CPD tasks are given.

\subsection{PRI Modulation Representation via Probabilistic Graphical Models}

\begin{figure}[!t]
\centering
\includegraphics[width=3.4in]{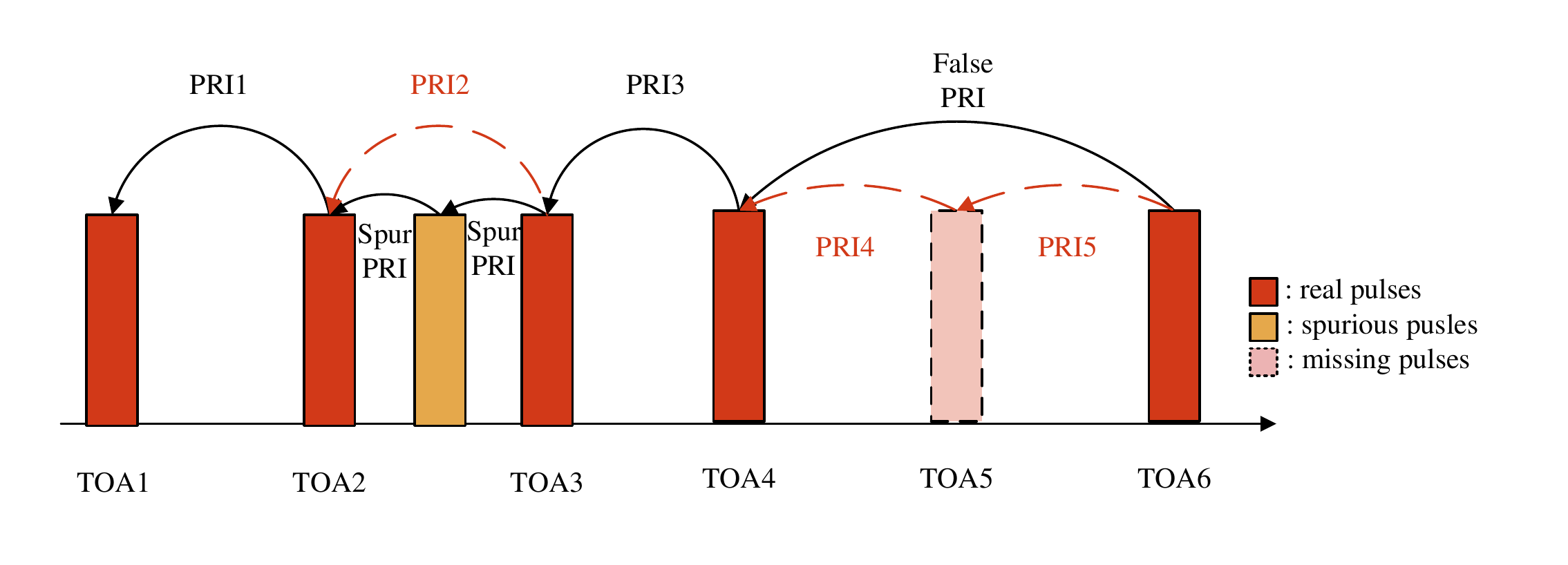}
\caption{Non-ideal observations in an MFR PRI sequence. The existence of a spurious pulse will lead original PRI value splits into two adjacent smaller $\Delta TOA$  values. A missing pulse will result in a large $\Delta TOA$  value, which represents
the summation of the original PRIs.}
\label{non-ideal_observations}
\end{figure}

There are six typical types of PRI modulations~\cite{wiley_elint_2006} that have been commonly used in related literature. Based on the hidden state self-transitioning tendency, common PRI types can be divided into two categories: agile and non-agile modulation types. Agile types including staggered PRI, sliding PRI, agile PRI. Non-agile types including jittered PRI, and Dwell and Switch (D\&S) PRI. Since PRI sequence is timing data and is extracted from Time Of Arrival (TOA) sequences through first-order difference. Non-ideal conditions caused by spurious pulses and missing pulses would affect the first-order difference value (such difference is denoted as $\Delta TOA$ for differentiation with the true PRI value) of its former pulses. The influence of a single missing pulse and a single spurious pulse in a pulse sequence are illustrated in Fig.~\ref{non-ideal_observations}. The existence of a spurious pulse will lead original PRI value splits into two adjacent smaller $\Delta TOA$  values. A missing pulse will result in a large $\Delta TOA$  value, whose value represents the sum of the PRI values of the missing pulse and it's previous pulse. To enhance the modeling capability under non-ideal conditions, this paper employs the probabilistic graphical models to represent various types of modulation types.


The MFR arranges a finite number of ordered pulses to achieve a certain radar work mode. Due to the measurement noise, the observation function of PRI values for each pulse can be modeled through a probability density function or probability mass function. This paper model the work mode as an HMM with Gaussian emissions (G-HMM). The G-HMM can be expressed as a three-tuple:
\begin{equation}
\label{equ_1}
    \boldsymbol{\lambda} = (\boldsymbol{\pi},\boldsymbol{A},\boldsymbol{\Theta})
\end{equation}
wherein $\boldsymbol{A}$ is the
state transition matrix with $K$ states; $\boldsymbol{\pi}$ is the initial state distribution, and $\boldsymbol{\Theta}$ is the set of Gaussian distributions. Following \textbf{Definition} defines the mapping between the G-HMM and different PRI defined work modes.

\begin{figure*}[!t]
\centering
\includegraphics[width=6in]{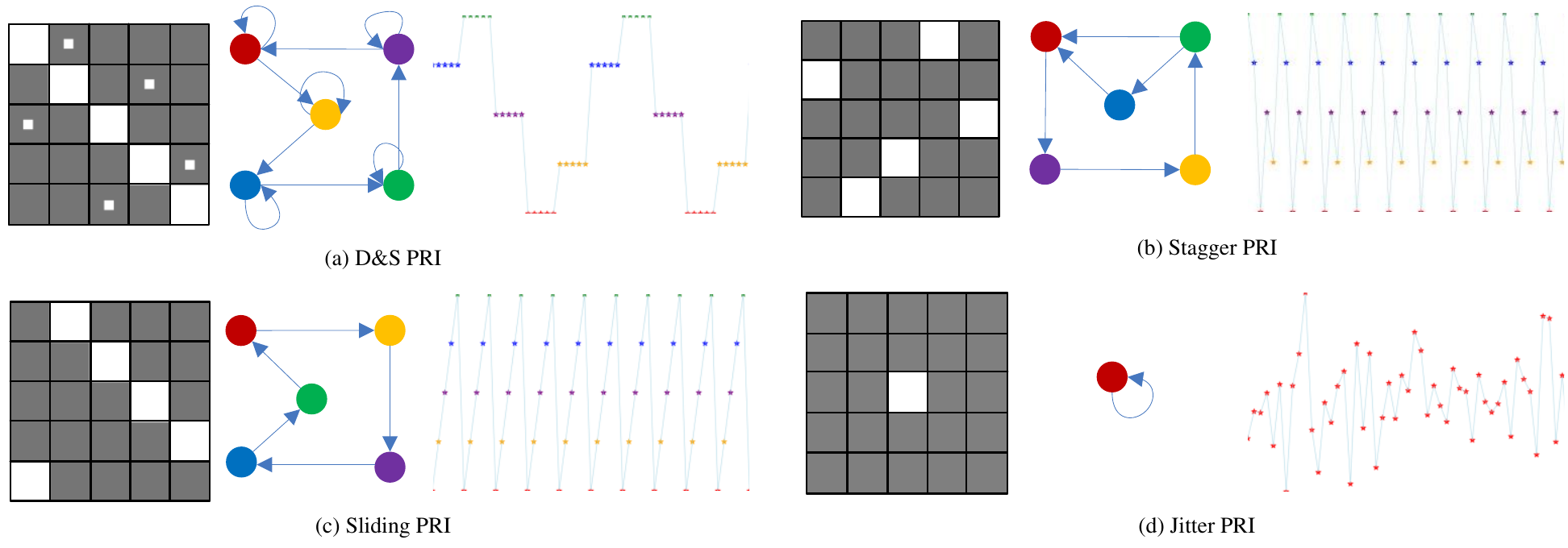}
\caption{PRI sequences and their corresponding Markov chain and state transition matrix. (a) D\&S PRI, (b) Stagger PRI, (c) Sliding PRI, (d) Jittered PRI.}
\label{exampleofpulsesequence}
\end{figure*}

\textbf{Definition}: The PRI sequence of a radar work mode can be represented by the Gaussian distribution set  $\boldsymbol{\Theta}=(\boldsymbol{\varphi}^1,...,\boldsymbol{\varphi}^k,...,\boldsymbol{\varphi}^K)$. In this study, $\boldsymbol{\varphi}^k$ represents the parametric model corresponds to the $k$th hidden state. Under Gaussian assumptions $\boldsymbol{\varphi}^k\sim N(\mu_k,\sigma_k^2)$. The probability density function (PDF) under $\boldsymbol{\varphi}^k$ is defined as:
\begin{equation}
\label{equ_2}
    f_{\boldsymbol{\varphi}^k}(p_t)=\frac{1}{\sqrt{2\pi\sigma_k^2}}\exp \bigg(-\frac{(p_t-\mu_k)^2}{2\sigma_k^2}\bigg),1\leq k \leq K
\end{equation}
wherein $p_t$ is the received pulse PRI value at time step $t$. The PRI sequence follows an underlying transition pattern $\boldsymbol{A}=(\boldsymbol{a}_j)_{j=1}^N,\boldsymbol{a}_j=(a_{ji})_{i=1}^N$. For instance, considering a work mode with staggered PRI modulation, the transmitted stagger sequence in a single period is $(2,3,5)$. In this case, the received stagger sequence with measurement noise (assume noise variance is 0.1) can be represented by $\boldsymbol{\Theta}=(\boldsymbol{\varphi}^1,\boldsymbol{\varphi}^2,\boldsymbol{\varphi}^3)$ with $\boldsymbol{\varphi}^1=(\mu_1=2,\sigma_1^2=0.1)$, $\boldsymbol{\varphi}^2=(\mu_2=3,\sigma_2^2=0.1)$, $\boldsymbol{\varphi}^3=(\mu_3=5,\sigma_3^2=0.1)$, respectively. Besides, $a_{12}=1, a_{23}=1, a_{31}=1, a_{ji}=0\, {\rm ~for~}\, ji\notin\{12,23,31\}$. Fig.~\ref{exampleofpulsesequence} shows an example of different modulated PRI sequences, corresponding graphical transitions and state transition matrix.

\subsection{MFR Work Mode Parameter Estimation and Change Point Detection}

We formulate the difference in either the inter-pulse modulation type or the modulation parameters as a change point in two adjacent work modes. Assuming $\{p_t\}_{t\geq1}$ is a pulse sequence parameterized by $\{\boldsymbol{\lambda}_t\}_{t\geq1}$. Until the change point $\nu$, the parameter stays at $ \boldsymbol{\lambda}_{<\nu}$, but after the change point, the parameter changes to $\boldsymbol{\lambda}_{\geq \nu}$, as shown in \eqref{equ_3}. 

\begin{equation}
\label{equ_3}
    \mathcal{H}(p_t,\boldsymbol{\lambda}_t)=
    \begin{cases}
    \boldsymbol{\lambda}_{<\nu} & \rm{if} \  \emph{t} < \nu\\
    \boldsymbol{\lambda}_{\geq \nu}& \rm{if} \  \emph{t} \geq \nu
    \end{cases}
\end{equation}



In traditional change point detection studies, the pre- and post-change distributions maybe known in advance (or at least their structures are known). But in radar interception applications with less prior assumption, we assume that the exact pre- and post-change distribution (both the structure and the parameters) is unknown. The structure of both distributions are modeled and determined by the proposed agile BNP-HMM, and the pre-change distribution can be estimated from the data according to the Bayesian theorem:
\begin{equation}
\label{equ_4}
    p(\boldsymbol{\lambda}_t|\boldsymbol{P}^t)=\frac{p(\boldsymbol{\lambda}_t)p(\boldsymbol{P}^t|\boldsymbol{\lambda}_t)}{\int p(\boldsymbol{P}^t,\boldsymbol{\lambda}_t)d\boldsymbol{\lambda}_t}
\end{equation}
wherein $p(\boldsymbol{\lambda}_t)$ is the prior distribution; $\boldsymbol{P}^t=\{p_t\}_{t=1}^t$ is a batch of data and $\int p(\boldsymbol{P}^t,\boldsymbol{\lambda}_t)d\boldsymbol{\lambda}_t$ is the evidence integral. Specifically,  the objective of the parameter estimation task is the number of PRI level in a period ($K$), the Gaussian parameters of each PRI level ($\boldsymbol{\varphi}^1,...,\boldsymbol{\varphi}^K$), and the modulation type of a pulse sequence ($\boldsymbol{A}$). After these hidden variables are inferred, a change point detection task will be performed through the estimated parameter sequence.

To clarify the detection criterion, the detector reaches the stopping time $N$ called the change point alarm time. From the online processing perspective, the alarm time should always happen after the actual change time. We consider Lorden's “worst case” Mean Detection Delay (MDD)~\cite{lorden_procedures_1971}:
\begin{equation}
\label{equ_5}
    \bar{\tau}^*=\sup _{t \geq 1} \operatorname{ess} \sup \mathbb{E}_{\boldsymbol{\lambda}_{\geq \nu}}\left(N-\nu \mid N \geq \nu, \boldsymbol{P}\right)
\end{equation}
The MDD should be as small as possible before a false alarm under the constraint of Mean Time to False Alarm (MT2FA), and the MT2FA is formulated as follows:
\begin{equation}
\label{equ_6}
    \overline{T}=\mathbb{E}_{\boldsymbol{\lambda}_{<\nu}}(N)
\end{equation}
wherein $\rm{ess} \sup$ denotes the essential supremum. The mean detection delay $\overline{\tau}^*$ needs to be minimized while the mean time to false alarm $\overline{T}$ needs to be maximized:
\begin{equation}
\label{object_CPD_task}
    \operatorname{minimize} \operatorname{MDD}(N) \text { subject to } \operatorname{MT2FA}(N) \geq \gamma
\end{equation}
wherein $\gamma$ is the constraint of MT2FA. The above optimization problem is a minimax problem.

\section{Method}
\label{sec:method}
This paper proposes an online processing framework for MFR work mode parameter estimation and change point detection. This section at first introduces a fully conjugate generative model (agile BNP-HMM) and a new prior distribution (the agile prior distribution). Then, we describe the online processing procedure and corresponding implementations.
\subsection{Agile Bayesian Non-parametric HMM Construction}
For non-cooperative tasks of MFR reconnaissance, the number of  hidden states and corresponding parameters of the work mode are always unknown and need to be inferred. A novel agile prior distribution is designed to control the G-HMM hidden state self-transitioning tendency to model the transition process of the MFR work mode with an agile inter-pulse modulation type. In this section, we first present a generative model for an MFR work mode. Then we introduce an improved stick-breaking construction method for the agile modulation type prior distribution implementation. Finally, we derived closed-form variational Bayesian iteration functions based on the proposed graphical model.

\subsubsection{Agile Bayesian Non-parametric HMM}
\begin{figure}[!t]
\centering
\includegraphics[width=3in]{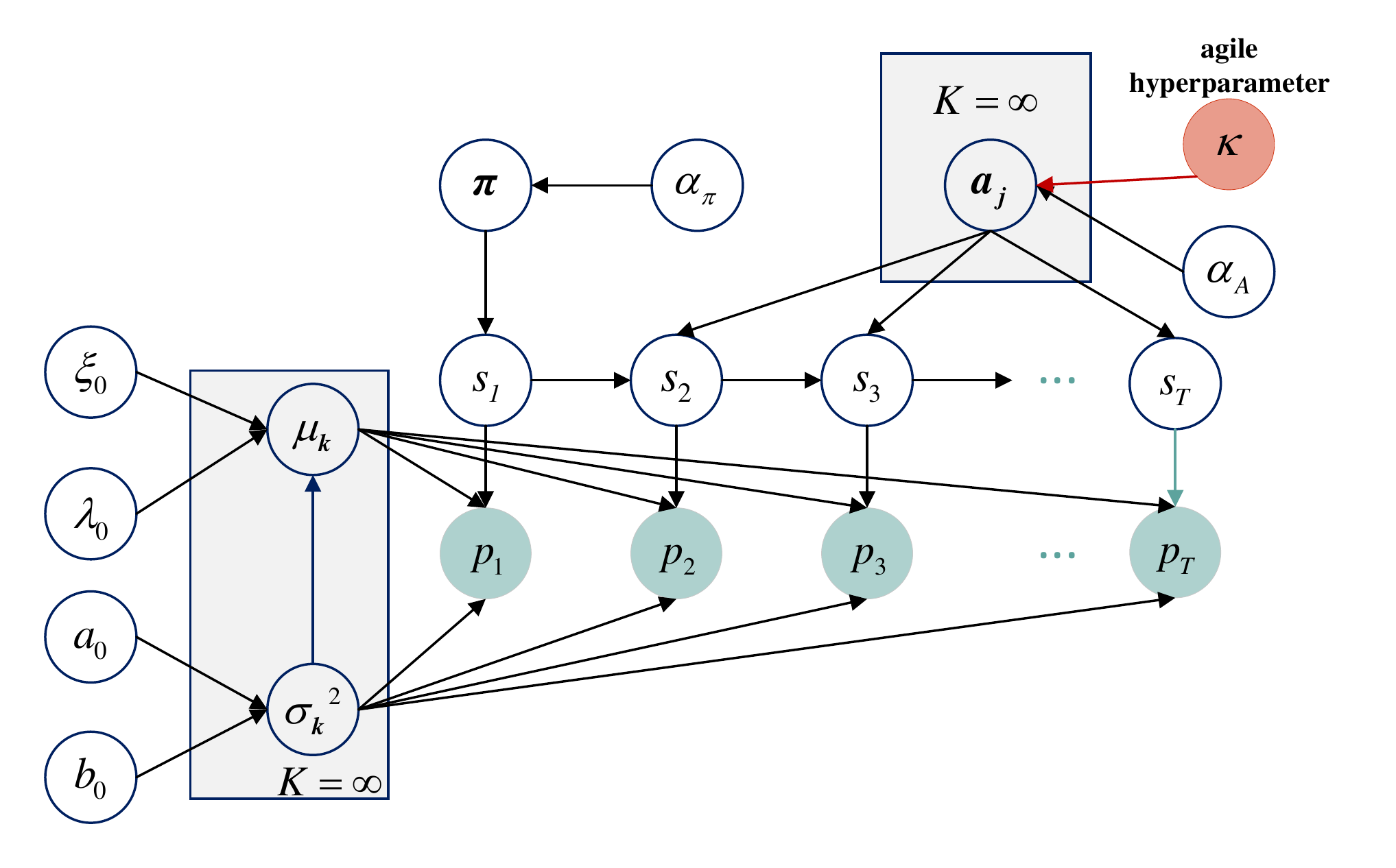}
\caption{The graphical model of the proposed agile BNP-HMM.}
\label{graphical_model}
\end{figure}

The graphical model of the  agile BNP-HMM is shown in Fig.~\ref{graphical_model}, wherein the intercepted pulses $\boldsymbol{P}=\{p_t\}_{t\geq1}$, hidden state assignments $\boldsymbol{S}=\{s_t\}_{t\geq1}$, sufficient statistics $\boldsymbol{\Theta} = ({\boldsymbol{\varphi}^k})_{k=1}^\infty$, initial distribution $\boldsymbol{\pi}=(\pi_k)_{k=1}^\infty$, transition matrix $\boldsymbol{A}=(\boldsymbol{a}_j)_{j=1}^\infty,\boldsymbol{a}_j=(a_{ji})_{i=1}^\infty$, concentration parameter $\alpha_\pi, \alpha_A$,  the agile hyper-parameter $\kappa$ and the length of the data batch $T$ are presented. The generative model can be formulated using the Bayesian theory as follows:

\begin{equation}
\label{equ_7}
    \begin{aligned}
     &p(\boldsymbol{P,S,\Theta,\pi,A}) \\
     &= p(\boldsymbol{P,S|\Theta,\pi,A})p(\boldsymbol{\pi})p(\boldsymbol{A})p(\boldsymbol{\Theta}) \\
    &= \Bigg[p(s_1|\boldsymbol{\pi})\prod_{t=2}^T p(s_t|s_{t-1},\boldsymbol{A})\prod_{t=1}^T p(p_t|s_t)\Bigg]\\
    &~~~~p(\boldsymbol{\pi})p(\boldsymbol{A})\prod_{k=1}^\infty p(\boldsymbol{\varphi}^k)
    \end{aligned}
\end{equation}

Based on the research presented in~\cite{bishop_pattern_2006}, to ensure the generative model is fully conjugate, the joint distribution of the sufficient statistics are generated using the Gaussian-Gamma distribution, which is given in~(\ref{equ_8}). 
\begin{equation}
\label{equ_8}
    \begin{aligned}
        p(\boldsymbol{\varphi}^k)&=N(\mu_k,\sigma_k^2) \\
        p(\mu_k|\sigma_k^{-2})&=N(\xi_0,\sigma_k^2/\lambda_0)\\
        p(\sigma_k^{-2})&=Gamma(a_0,b_0)
    \end{aligned}
\end{equation}
wherein $\mu_k,\sigma_k^2$ are the mean and variance of the Gaussian distribution, respectively; $\sigma_k^{-2} =\frac{1}{\sigma_k^2}$ represents the precision of the Gaussian distribution; $\xi_0,\lambda_0,a_0,b_0$ are hyper-parameters of the Gaussian-Gamma distribution; $N$ refers to Gaussian distribution and $Gamma$ refers to Gamma distribution\footnote{In MFR applications, each work mode is modeled by a BNP-HMM. The hidden state number in BNP-HMM generally varies across different MFR work modes. The work mode can change over time, and the number of hidden states would also change over time. The infinite assumption in 
Bayesian-nonparametric theory does not require the number of hidden states be pre-defined during inference}.



The Dirichlet distribution to the HMM encourages hidden states to have similar transition distributions. However, it does not differentiate self-transitions from moves between the hidden states. Therefore, when modeling the MFR work mode with the agility characteristic, the self-transition nature of the Dirichlet distribution allows for a high posterior probability of hidden state persistence. Thus it tends to underestimate the hidden state number.

This paper proposed the agile BNP-HMM to determine the number of hidden states. The agile BNP-HMM is governed by placing an agile prior distribution over the state transition probability. The agile prior distribution is used to present the case of a few self-transitions. To implement the agile prior distribution, a novel stick-breaking construction method is designed.
\subsubsection{Agile Prior and Stick-breaking Construction}
At first, the Dirichlet distribution is reviewed. For any partitions  $G=\{G_1,...,G_\infty\}$ of the probability space $\Omega$, the distribution of the base measure's probability mass on the partition $G$ is defined as follows:
\begin{equation}
\label{equ_9}
\begin{aligned}
    &(a_{j1},a_{j2},...,a_{j\infty})=\\
    &\mathcal{DIR}(\alpha_A H(G_1),...,\alpha_A H(G_\infty))
\end{aligned}
\end{equation}
wherein $\mathcal{DIR}$ denotes the Dirichlet process; $H$ is the base measure; $\alpha_A$ represents the concentration parameter; $\boldsymbol{a}_j$ is a random variable that follows the multi-nominal distribution, and the $\mathcal{DIR}$ can be implemented as a stick-breaking construction\footnote{The construction can be briefly explained. Suppose there is a stick with length 1. Let $a_{ji}' \sim Beta(1,\alpha_A)$ for $i=1,2,3,...,\infty$, and regard them as fractions for how much we take away from the remainder of the stick every time. Then $a_{ji}$ can be calculated by the length we take away each time: $a_{j1}=a_{j1}',a_{j2}=(1-a_{j1}'),...,a_{ji}=a_{ji}'\prod_{n=1}^{i-1}(1-a_{jn}')$. By the stick-breaking construction, we will have $\sum_{i=1}^\infty a_{ji}=1$.}~\cite{sethuraman1994constructive}:
\begin{equation}
    \begin{aligned}
        a_{ji}'&= Beta(1,\alpha_A), \\ 
        a_{ji}&=a_{ji}'\prod_{n=1}^{i-1}(1-a_{jn}')
    \end{aligned}
\end{equation}
wherein $Beta$ refers to the beta distribution and $\alpha_A$ is the concentration parameter.


An agile hyper-parameter is introduced to the Dirichlet distribution to control the hidden states self-transitioning tendency of MFR pulse sequences, resulting in agile Dirichlet process (agile $\mathcal{DIR}$), which is given by:
\begin{equation}
\begin{aligned}
    &(a_{j1},...,a_{jj},...,a_{j\infty})=\\
    &\mathcal{DIR}(\alpha_A H(G_1),...,(\alpha_A+\kappa)H(G_j),..., \alpha_A H(G_\infty))
\end{aligned}
\end{equation}
wherein $\kappa$ is the agile hyper-parameter, and a larger $\kappa$ value encourages an HMM to move from the current state $j$ to another state instead of staying in the current state. 

The agile $\mathcal{DIR}$ can be implemented via a new stick-breaking construction as follows:
\begin{equation}
\label{equ_10}
    \begin{aligned}
        a_{ji}'&= Beta(1,\alpha_A+\kappa \delta(i-j)), \\ 
        a_{ji}&=a_{ji}'\prod_{n=1}^{i-1}(1-a_{jn}')
    \end{aligned}
\end{equation}
wherein $\sum_i^\infty a_{ji}=1$, and $\delta$ is the Dirichlet-delta function. When $i=j$ , $a_{jj}'$ is sampled from $Beta(1,\alpha_A+\kappa)$, the $Beta$ distribution tends to put more mass on smaller values. Thus $a_{jj}'$ tends to smaller than $a_{ji}'$, leading to break a shorter stick.

\subsubsection{Variational Inference for Agile BNP-HMM}

In the inference procedure, it is necessary to estimate the posterior distribution from the observed data and the given hyper-parameters $\boldsymbol{\Upsilon}=\{\alpha_\pi,\alpha_A,a_0,b_0,\lambda_0,\xi_0,\kappa\}$. The inference of agile BNP-HMM is illustrated as Algorithm \ref{agileBNPHMM}. 


Variational inference is an efficient way to approximate the intractable posterior, which introduces a variational distribution \textit{q} to approximate the true posterior. The basic idea of variational inference is to minimize the distance between the variational distribution \textit{q} and the exact posterior \textit{p}. The likelihood function is formulated as \eqref{equ_12}:
\begin{equation}
\label{equ_12}
\begin{aligned}
    &\ln p(\boldsymbol{P|S,\pi,A,\Theta,\Upsilon}) = \mathcal{L}(q(\boldsymbol{S,\pi, A,\Theta}))\\
    &+ \mathcal{KL}(q(\boldsymbol{S,\pi,A,\Theta})||p(\boldsymbol{S,\pi,A,\Theta|P,\Upsilon}))
\end{aligned}
\end{equation}
wherein $\mathcal{L}$ denotes the well-known evidence lower bound (ELBO), and $\mathcal{KL}$ is the Kullback-Leibler divergence. Minimizing the similarity between the variational distribution and the exact posterior is to maximize the evidence lower bound~\cite{bishop_pattern_2006}. 

This paper considers the partial mean-field variational family \cite{ghahramani1995factorial} and truncation assumption for variational inference. The partial mean-field family assumes that $(\boldsymbol{\pi,A,\Theta,\Upsilon})$ and $(\boldsymbol{S})$ are mutually independent. The truncation assumption considers the probabilities of $L$ states as the infinite hidden states, wherein $L$ is called the truncation level, and $L$ should be large enough to ensure the accuracy. The variational distribution is defined as follows:
\begin{equation}
\label{equ_13}
    \begin{aligned}
        q(\boldsymbol{\Theta})&=q(\boldsymbol{S})q(\boldsymbol{\pi}')q(\boldsymbol{A}')q(\mu,\sigma^{-2}) \\
        &=q(s_1)\prod_{t=2}^Tq(s_t|s_{t-1})\prod_{i=1}^Lq(\pi_i')\prod_{j=1}^L \prod_{i=1}^L q(a_{ji}')\\
        &~~~~\prod_{j=1}^L q(\mu_j|\sigma_j^{-2})q(\sigma_j^{-2})
    \end{aligned}
\end{equation}

Exponential family is chosen for each component of the variational distribution. The variational distributions can then be represented as follows: 
\begin{equation}
\label{equ_14_1}
        q(\pi_i')=Beta(\eta_1(\pi_i'),\eta_2(\pi_i'))
\end{equation}
\begin{equation}
    \label{equ_14_2}
        q(a_{ji}')=Beta(\eta_1(a_{ji}'),\eta_2(a_{ji}'))
\end{equation}
\begin{equation}
    \label{equ_14_3}
        q(\mu_j|\sigma_j^{-2})=N(\mu_j,\sigma_j^{-2}/\lambda_j)
\end{equation}
\begin{equation}
    \label{equ_14_4}
        q(\sigma_j^{-2})=Gamma(a_j,b_j)
\end{equation}

Then the evidence lower bound can be derived according to:
\begin{equation}
\label{equ_15}
     \mathcal{L} = \mathbb{E}_q[\ln p(\boldsymbol{\pi,S,A,\Theta,P})]-\mathbb{E}_q[\ln q(\boldsymbol{\pi,S,A,\Theta})]
\end{equation}

This paper uses the coordinate ascent algorithm~\cite{blei_variational_2017} to maximize the ELBO. The universal update function is given by \cite{bishop_pattern_2006}:
\begin{equation}
    \label{universialVI}
\ln \left[q_n^*\left(\boldsymbol{Z}_n\right)\right]=\mathbb{E}_{-n}[\ln p(\boldsymbol{P}, \boldsymbol{Z})]+\text { const }
\end{equation}
wherein $\boldsymbol{Z}=\{ \boldsymbol{\pi,A,\Theta,S}\}$, $n$ refers to the $n$th hidden variable. The details of the derivation and the update function can be found in the supplementary material.

\begin{algorithm}
\label{agileBNPHMM}
  \caption{Agile Bayesian Non-parametric HMM (agile BNP-HMM)}\label{alg:alg1}
  \KwIn{a batch of data $\boldsymbol{P}^n$, hyperparameters $\boldsymbol{\Upsilon}$, truncation level $L$, and prior distribution $prior$}
  \KwOut{transition matrix $\boldsymbol{A}$, parameter set $\boldsymbol{\Theta}$, and initial distribution $\boldsymbol{\pi}$}
    \eIf{prior is None}{
        initialize $\boldsymbol{\pi},\boldsymbol{A},\boldsymbol{\Theta},\boldsymbol{S}$\;
    }{
        $\boldsymbol{\pi,A,\Theta,\boldsymbol{S}}\leftarrow prior$\;
    }
    \Repeat{
           Convergence
        }{
        Variational Inference \eqref{universialVI}\;
        }
    Sample $\boldsymbol{\pi,A,\Theta}$ and return\;
\end{algorithm}

\subsection{Implementation of MFR Work Mode Parameter Estimation and Change Point Detection}
This paper considers a control chart base upon the weighted SPRT for online change point detection paradigm. To facilitate the inference in each new time step, streaming Bayesian updating\cite{broderick_streaming_2013} is designed to the iterative variational inference of the agile BNP-HMM model. This streaming process also prevents variational inference from falling into a local optimal.

\subsubsection{Initialization of the Agile BNP-HMM-CUSUM Framework}
After setting the hyper-parameters $\Upsilon$, pulse PRI values were collected in a buffer called the initial-batch. Since this paper assumes no knowledge about the data in the initial-batch, the Dirichlet Process Mixture Model (DPMM)~\cite{blei_variational_2006} is used to obtain an initial guess of agile BNP-HMM model parameters of the data in initial-batch. For subsequent PE task, due to the streaming nature, six hyper-parameters that need to be tuned $\boldsymbol{\Upsilon}=\{\alpha_A,\alpha_\pi,\xi_0,\lambda_0,a_0,b_0\}$, wherein $\boldsymbol{\Upsilon}$ is time variant and thus not suitable for change point detection tasks. In this work, the agile hyper-parameter $\kappa$ represents the time-series characteristics over time, and $\kappa$ is the only parameter need to be tuned for the PE task. The whole framework is illustrated as Algorithm \ref{agileBNPHMMCUSUM}.

Since our model is fully conjugate, we only need to initialize the sufficient statistics of \eqref{equ_14_1}-\eqref{equ_14_4}. Specifically, $\eta_1(\pi_1'),\eta_2(\pi_2'),\eta_1(a_{ji}'),\eta_2(a_{ji}'), a_j,b_j$ are all set to one in the initialization step. The mean $\mu_j$ and variance $\sigma_j^{-2}/\lambda_j$ of each Gaussian distributions are set to 0 and 0.1 in the initialization step.

\subsubsection{Online Parameter Estimation Task for Agile BNP-HMM under Non-ideal Observations}
After initialization, each pulse in the initial-batch belongs to a particular hidden state. In parameter estimation step, a new pulse is received and combined with previously arrived pulses to create a new data batch. Given the new data batch, the agile BNP-HMM algorithm is used to update the transition matrix and  Gaussian distribution parameters. Due to the non-ideal observation conditions, there will be $\Delta TOA$ values caused by missing and spurious pulses. These $\Delta TOA$ values will result in a false hidden state with a relatively low visiting probability and a high transition probability. An acceptance threshold $\gamma_u$ is set, when $\mathbb{E}[\boldsymbol{a}_j]\leq \gamma_u$ satisfies, the corresponding hidden states are considered as $\Delta TOA$ values caused by missing or spurious pulses and are deleted. The remaining hidden states are assumed originating from true PRI values.

Streaming Bayesian updating enables high efficient parameter estimation and reduces the risk of falling into local optima during the PE task. Assume that $\boldsymbol{P}^n=\{p_t\}_1^{m+n}$, and index $m$ and $n$ as the initial-batch length and the $n$th arrived pulse after the initial-batch, respectively. For the initial-batch, applying the Bayesian rule yields $p(\boldsymbol{Z}^0|\boldsymbol{P}^0)=p(\boldsymbol{P}^0|\boldsymbol{Z}^0)/p(\boldsymbol{P}^0)$. Superscript $0$ denotes batch index $0$. When a new pulse arrives, the batch index increases by one and a random hidden variable is assigned to the new arrived pulse. All underlying state sequences are combined and denoted as $p(\boldsymbol{Z}^1)$. By applying the Bayesian rule again, instead of placing the prior $p(\boldsymbol{Z}^1)$, the previous data batch’s variational posterior $q(\boldsymbol{Z}^0|\boldsymbol{P}^0)$ is set as the current data batch’s prior. In variational inference, the true posterior for the $n$th batch $p(\boldsymbol{Z}^n|\boldsymbol{P}^{1:n})$ is approximated by the variational distribution $q^n(\boldsymbol{Z}^n)$. Thus, the online variational inference is obtained by substituting $q^{n-1}(\boldsymbol{Z}^{n-1})$ for the prior as follows:
\begin{equation}
    p\left(\boldsymbol{Z}^{\mathrm{n}} \mid \boldsymbol{P}^{1: n}\right) \approx q^n(\boldsymbol{Z})=\frac{p\left(\boldsymbol{P}^n \mid \boldsymbol{Z}^{\mathrm{n}}\right) q^{n-1}\left(\boldsymbol{Z}^{\mathrm{n}-1}\right)}{p\left(\boldsymbol{P}^{1: n-1}\right)}
\end{equation}

The estimated parameter tuple $(\boldsymbol{\pi},\boldsymbol{A},\boldsymbol{\Theta})^n$ for the $n$th batch is sent to a stack. All parameter tuples estimated in the previous time step are pushed in the stack. Then, the newly estimated tuple and the stack are sent to the change point detector.

\subsubsection{Online Change Point Detection Strategy}
Detecting the change point of the MFR work mode is equivalent to detecting the change point for BNP-HMM. However, BNP-HMM's non-ergodicity and infinite hidden state space render general results \cite{fuh_sprt_2003,fuh_asymptotically_2021} invalid. Additionally, computing the invariant measure of HMM is computationally demanding. Directly detecting changes from HMM contradicts the basic principle of designing a system in MFR applications - that it should be computationally efficient while clear enough to distinguish various work modes. Therefore, we treat the prior distribution and the transition distributions as nuisance parameters similar to \cite{fuh_quickest_2015}, which makes previous research results \cite{nikiforov_quadratic_1999} valid. Our approach resembles Stephen Boyd’s method \cite{hallac2017toeplitz}  by transforming learning complex structure problem into multivariate Gaussian inverse covariance matrices to simplify computationally intensive problems.

The change point detector considers the weighted SPRT~\cite{nikiforov_quadratic_1999}. Taking modulation parameter level change point as an example, the agile BNP-HMM sequentially returns the parameter set  $\boldsymbol{\Theta}_0,\boldsymbol{\Theta}_1,\boldsymbol{\Theta}_2,...,\boldsymbol{\Theta}_{L-m}$. The parameter sets can be described as a multi-variate Gaussian distribution as follows:
\begin{equation}
\label{equ_16}
    \mathcal{H}(\boldsymbol{\Theta}_t)=
    \begin{cases}
    N(\boldsymbol{\theta}_0,\boldsymbol{\Sigma}) & \rm{if} \  t < \nu\\
    N(\boldsymbol{\theta}_1,\boldsymbol{\Sigma}) & \rm{if} \  t \geq \nu
    \end{cases}
\end{equation}
wherein $\boldsymbol{\theta}_0=(\boldsymbol{\mu}_0^1,...,\boldsymbol{\mu}_0^K)$ and $\boldsymbol{\theta}_1=(\boldsymbol{\mu}_1^1,...,\boldsymbol{\mu}_1^K)$ are the pre- and post-change parameters, respectively. $\boldsymbol{\Sigma}$ is the unit covariance matrix. 

Following common assumptions in asymptotic performance analysis of the change point detector, we assume that the pre-change parameter $\boldsymbol{\theta}_0$ is known, and the post-change parameter $\boldsymbol{\theta}_1$ is unknown. The change amplitude $b$ is known and given by:
\begin{equation}
    b^2=\boldsymbol{(\theta}_1-\boldsymbol{\theta}_0)^{\rm{T}}\boldsymbol{\Sigma}^{-1}\boldsymbol{(\theta}_1-\boldsymbol{\theta}_0)
\end{equation}
The stopping time $N$ is expressed by:
\begin{equation}
\label{CUSUM}
    N=\inf \left\{t \geq 1: \max _{1<k<t-m+1} D_k^t>\lambda\right\}
\end{equation}
\begin{equation}
\label{statistic}
    D_k^t=-\frac{(t-k) b^2}{2}+\ln G\left(\frac{K}{2}, \frac{b^2(t-k)^2\left(\chi_k^t\right)^2}{4}\right) 
\end{equation}
\begin{equation}
\label{chi_2}
    \left(\chi_k^t\right)^2=\left(\bar{\boldsymbol{S}_k^t}\right)^T \boldsymbol{\Sigma}^{-1} \bar{\boldsymbol{S}}_k^t
\end{equation}
wherein $\bar{S}_k^t=\sum_{i=k}^t\left(\boldsymbol{\Theta}_i-\boldsymbol{\theta}_0\right)$,$G(d, z)=1+\frac{z}{d}+\cdots+\frac{z^n}{d(d+1) \ldots(d+n-1) n !}+\cdots$ is generalized hyper-geometric function, wherein $n$ is the index of the $n$ th item of the expansion. The strategy is asymptotic optimal in the sense of mini-max criterion~\cite{nikiforov_quadratic_1999}. The result is stand by the following theorem:
\newtheorem{theorem}{\bf Theorem}
\begin{theorem}\label{thm1}
The “worst-case” mean detection delay for the $\chi^2$-CUSUM is given by the following asymptotic equation as $\bar{T}\to \infty$.

\begin{equation}
\label{asymptotic}
    \bar{\tau}^*(\bar{T}) \sim \frac{\ln \bar{T}}{\rho\left(\boldsymbol{\theta}_1, \boldsymbol{\theta}_0\right)} =\frac{2 \ln \bar{T}}{b^2}
\end{equation}
wherein $\bar{\tau}^*$ is the optimal mean detection delay; $\bar{T}$ is the mean time to false alarm; $\rho=\frac{b^2}{2}$ is the Kullback-Leibler divergence of pre- and post-change distribution. 
\end{theorem}

\begin{figure}[!t]
\centering
\includegraphics[width=3.3in]{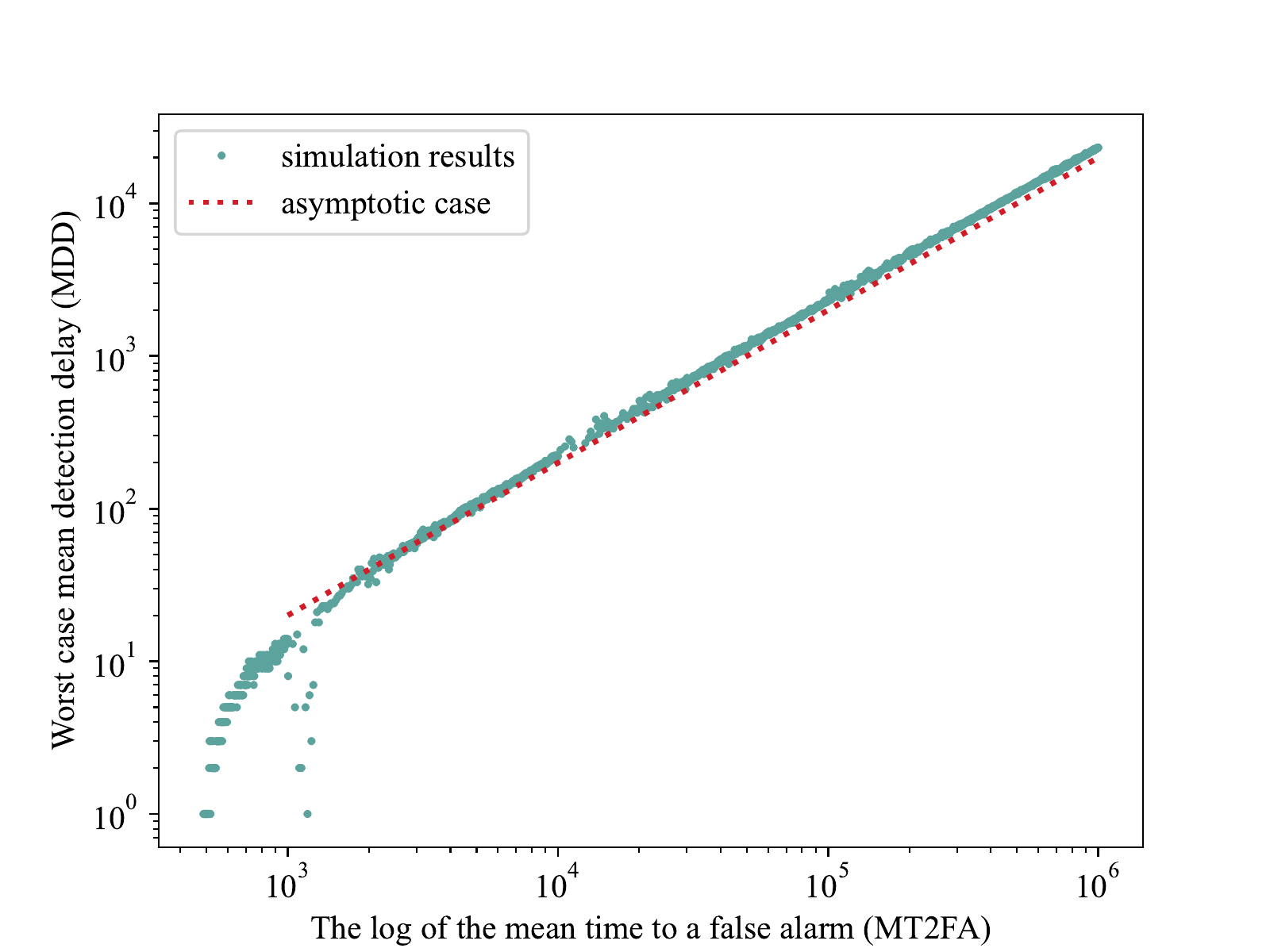}
\caption{Comparison between the simulation results and the asymptotic case.}
\label{asymptotic_curve}
\end{figure}

The proof of Theorem 1 can be found in supplementary material\footnote{Currently, we do not consider the situation of changes only in transition matrix between adjacent work modes. Because this situation is relatively rare in MFR applications. But as stated in the Introduction Section, the change point detection of this situation is non-trivial. We leave this for future study.}. We also show that the result of the comparison between the simulation results (green dots) and the asymptotic case (red dashed line) in Fig.~\ref{asymptotic_curve}, wherein each green dot represents the average value of 100 Monte Carlo simulations. The linear relationship between the MDD and the log of MT2FA in asymptotic cases is verified. The simulation results approach the asymptotic case when $\ln{\bar{T}}$ is large (such as $\ln{\bar{T}}\geq3 \times10^3$). When $\ln \bar{T}\to \infty$ is not satisfied, the Berk's theorem is not satisfied (see supplementary material), and the linear relationship between the $\ln \bar{T}$ and $\bar{\tau}^*$ can not be guaranteed. Meanwhile, a smaller value of $\ln \bar{T}$ implies that it is easier for the detector to trigger an alarm. Consequently, a smaller value of $\ln \bar{T}$ leads to a smaller value of MDD. This explanation also clarifies why the MDD of the simulation result (green dots) is lower than the asymptotic case (red dashed line) when the value of $\ln \bar{T} \to \infty$ is small.

\subsection{Complexity Analysis}

According to Blei~\cite{blei_variational_2017}, the complexity of the designed variational family determines the complexity of the optimization. The agile BNP-HMM chooses the structured (partial mean field) variational family~ \cite{ghahramani1995factorial}. The time complexity of one single iteration is $O(L^2T)$ and the space complexity is $O(LT)$, wherein $L$ is the truncation level and $T$ is the length of the sequence. The total time complexity of the algorithm is $O(NL^2T)$ (in space $O(LT)$), wherein $N$ is the number of iterations. In this paper, we have designed an online variational inference method under the variational streaming Bayes framework~\cite{broderick_streaming_2013}.  Under streaming settings, the PE task is expected to converge in fewer steps compared to random initialization or initiated by fixed values, resulting in a lower $N$. Meanwhile the PE task did not need to revisit the old data for current inference, resulting in a lower $T$. Thus, the computational redundancy can be reduced.

\subsection{Error Analysis for Non-ideal Conditions}
In this section, we establish a lower bound on the error probability for distinguishing true pulses from spurious ones using PRI data and maximum likelihood criterion. This bound can be used to assess how close an algorithm is to optimum, regardless of the specific algorithm. Although our focus is on spurious pulses, this bound can also be extended to cases involving missing pulses or mixtures of missing and spurious pulses\footnote{The derivation is inspired by \cite{young_deinterleaving_2018}, which establish a lower bound for the deinterleaving scheme of mixtures of renewal process.}.

In general, estimation error of the PRI value of a true pulse as a spurious pulse can significantly impact the estimation of subsequent modulation types and parameters. Specifically, defining the $m$th pulse as the true pulse, and the $m'$th pulse be the nearest spurious pulse. The pulse index $m'$ is formulated by $m'={\rm arg}\min_{m'}|TOA_m-TOA_{m'}|, m\in \Omega_t, m'\in\Omega_s$, wherein $TOA_m$ is the arrival time of the $m$th pulse, the $\Omega_t$ is the set of true pulse's indices, and $\Omega_s$ is the set of spurious pulse's indices. Let $M_t$ represent the event wherein the $m$th true pulse is identified as a spurious pulse and $m'$th spurious pulse is identified as a true pulse. We assess the performance of our algorithms by the error probability\footnote{Other possible errors are not considered in this study as we are developing a lower bound.}: 
\begin{equation}
\label{errorprobability}
P_e = \lim_{T\to\infty} P(\hat{s_j}\neq s_j)
\end{equation}
wherein $s_j$ is the true hidden state, and $\hat{s_j}$ is the predicted hidden state. Then the lower bound on the error probability of the true pulse therefore is:
\begin{equation}
\label{lowerbound}
\begin{aligned}
P_e^t &\geq \lim _{T \rightarrow \infty} E\left[\frac{1}{T} \sum_{t=1}^T \mathds{1}\left(M_t\right)\right]=\lim_{T \rightarrow \infty} \sum_{t=1}^T P(M_t)=P\left(M_t\right)
\end{aligned}
\end{equation}
wherein $T$ is the number of received pulses, $\mathds{1}$ is the indicator function. To derive the lower bound, we make two assumptions. Firstly, the elements of event $\{M_t\}_{t=1}^T $ are disjoint. Secondly, we assume that the true pulses and the spurious pulses are mutually independent. Theorem 2 provides the error probability lower bound.
\begin{theorem}\label{thm2}
Suppose the true pulse and the spurious pulse can be represented by $\varphi_t\sim N(\mu_t,\sigma_t^2 ),\varphi_s\sim N(\mu_s,\sigma_s^2)$, respectively. The lower bound on error probability for the true pulse is:
\begin{equation}
\label{lowerbound}
\begin{aligned}
&P_e^t \geq \\
&\int_0^{\infty} \int_{\frac{\sigma_s}{\sigma_t} x}^{\infty} \frac{\exp \left(-\frac{k^2}{4}\right)}{2 \sqrt{\pi} Q\left(\frac{\sigma_s}{\sigma_t} x-\frac{\mu_t}{\sigma_t}\right) Q\left(2 \frac{\sigma_s}{\sigma_t} x-\frac{\mu_t}{\sigma_t}\right)}\\
&\bigg(Q\left[\sqrt{2}\left(\frac{\sigma_s}{\sigma_t} x\right.\right.
\left.\left.-\frac{\mu_t}{\sigma_t}-\frac{k}{2}\right)\right]+Q\left[\sqrt{2}\left(2 \frac{\sigma_s}{\sigma_t} x-\frac{\mu_t}{\sigma_t}-\frac{k}{2}\right)\right]\\
& -1\bigg) dk \int_{-\frac{x}{2}}^{\infty} \frac{1}{Q\left(-\frac{x}{2}\right)\sqrt{2 \pi}} \exp \left(-\frac{k^2}{2}\right) d k\\
&\frac{\sigma_s}{\mu_s} Q\left(x-\frac{\mu_s}{\sigma_s}\right) d x
\end{aligned}
\end{equation}
wherein $Q(x)$ is the right tailed function. 
\end{theorem}

The proof can be found in the supplementary meterial. We also compare the performance of agile BNP-HMM, Viterbi algorithm and Maximum A Posterior (MAP) algorithm \cite{bishop_pattern_2006} to predict the membership of each pulse and the lower bound of Theorem 2. The error probability for finite time on their outputs as:
\begin{equation}
    \label{errorprob}
    P_e^t=\frac{\sum_i \mathds{1}\left(\hat{s}_i \neq s_i,s_i=t\right)}{\sum_i \mathds{1}\left(s_i=t\right)}
\end{equation}

\begin{figure}[!t]
\centering
\includegraphics[width=3.3in]{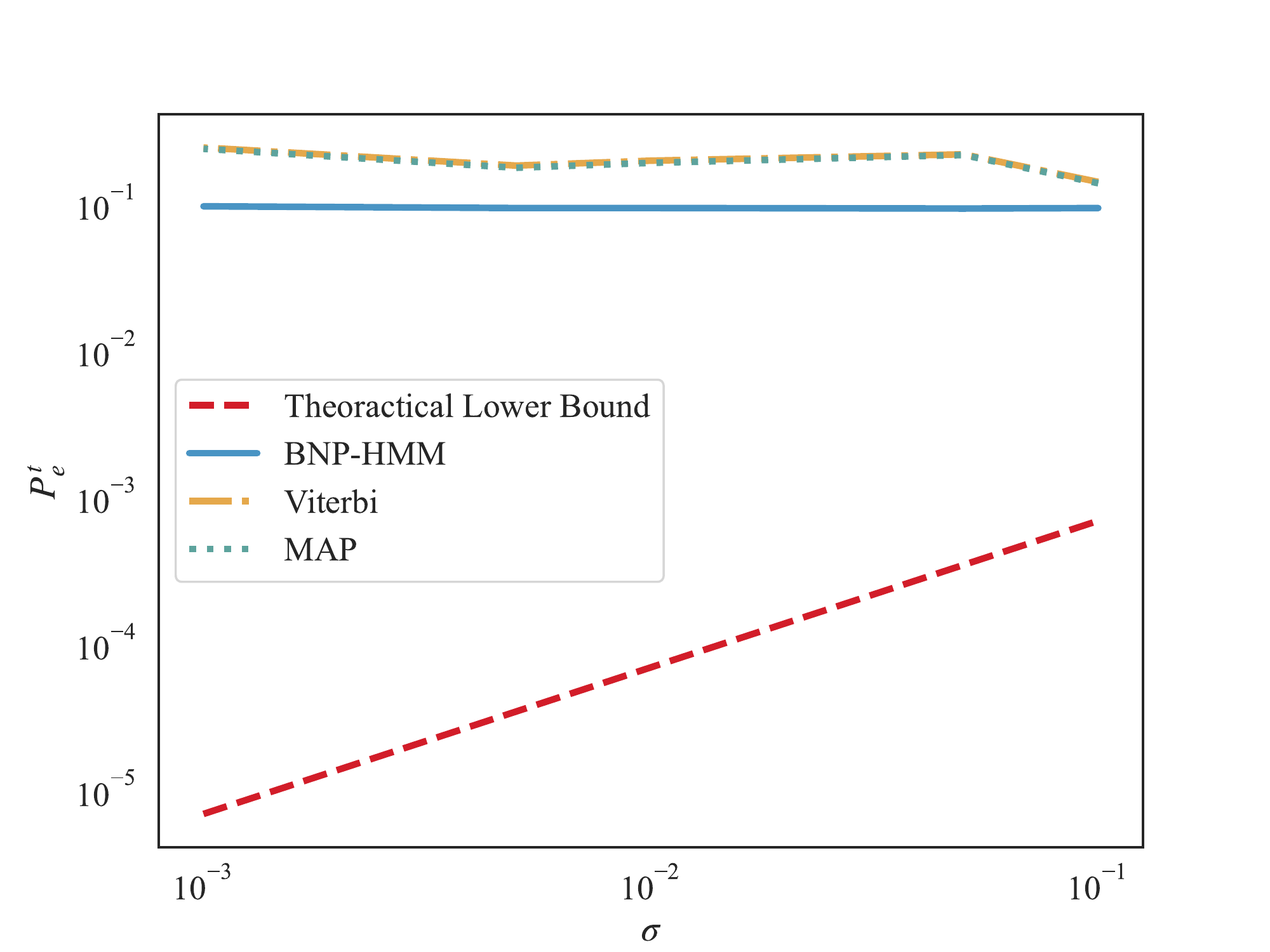}
\caption{The error probabilities in the true pulse, obtained from the theoretical lower bound and the BNP-HMM, Viterbi, and MAP algorithms, were compared.}
\label{theoracticalbound}
\end{figure}

Set the true pulses follow $\varphi_t\sim N(3,\sigma_t^2 )$, and the spurious pulses follow $\varphi_s\sim N(1.5,\sigma_s^2 )$, and let $\sigma_t=\sigma_s=\sigma$. We varied the standard deviation from $10^{-3}$ to $10^{-1}$ and the result is the average of 100 times Monte Carlo simulations. As shown in Fig.~\ref{theoracticalbound}, the results indicate that the BNP-HMM is closer to the lower bound than the conventional Viterbi and the MAP algorithm.

\section{Simulated Performance on Radar Data}
\label{sec:simulated performance on radar data}
Simulations of the PE task and the CPD task are conducted using PRI-defined sequences to examine the effectiveness and robustness of the proposed method. The datasets and evaluation metrics are described in Section~\ref{subsec:simulation design}, and simulation results and detailed discussions are presented in Section~\ref{subsec:PEtask} to Section~\ref{subsec:CPDtask}.

\subsection{Simulation Design}
\label{subsec:simulation design}
\subsubsection{Simulation Settings}
Simulation datasets included PRI sequences with multiple segments generated by the G-HMM described in Section~\ref{sec:method}. Given the modulation type, the transition matrix of the single Work Mode (WM) is randomly sampled, and corresponding parameters are sampled based on the G-HMM sufficient statistics. The modulation types and parameters of each simulation category are given in Table \ref{datasets}. Square brackets contain modulation parameters. The three elements in the parentheses of sliding PRI and D\&S PRI indicate the minimum and maximum values for sliding PRI, as well as the number of PRI levels within a period. The number of one PRI level of D\&S PRI is randomly sampled from 5 to 10.  Jittered PRI is represented by parentheses indicating its mean value and variance.  Considering the signal propagation error and the measurement error, additive Gaussian measuring noise with zero mean and variance 1 is added.

The first-category of the simulation explored the PE performance of agile BNP-HMM for different agile hyper-parameter $\kappa$ values. Each pulse sequence contains pulses from one single work mode with 1000 pulses. The comparison of results and the discussions are given in Section~\ref{subsec:PEtask}.

The second and the third categories of simulation examine the CPD performance, including the detection accuracy and timing performance. Each pulse sequence sample contains four work modes, each of which lasts 150 pulses. The work mode change in the Same modulation Type and Variable modulation Parameters (STVP) (D2–D5), and Variable modulation Types and Variable modulation Parameters (VTVP) (D6) are simulated in the second and the third categories, respectively. The results of the STVP, VTVP and different parameter settings are presented in Section~\ref{subsec:CPDtask}.

Non-ideal conditions are added in each datasets.  A length $N$ pulse sequence with $n$\% non-ideal conditions is consisted with $N\times n/2\%$ number of randomly generated spurious pulses and $N\times n/2\%$  randomly generated missing pulses, respectively.

\begin{algorithm}
\label{agileBNPHMMCUSUM}
  \caption{Change point detection based on the $\chi^2-$CUSUM strategy and online Bayesian updating}\label{alg:alg2}
  \KwIn{ initial-batch $\boldsymbol{P}^0$, sequentially arrived pulses $\{p_t\}_{m+1}^{m+n}$, threshold $\gamma_p$, acceptance threshold $\gamma_u$.}
  \KwOut{change point index $\boldsymbol{d}$}
  Initialize statistic of CPD task: $D_p=0$\;
  Initialize indicator variable $\boldsymbol{d}$\;
  $\boldsymbol{\pi}^0,\boldsymbol{A}^0,\boldsymbol{\Theta}^0$ $\leftarrow$ agile BNP-HMM($\boldsymbol{P}^0$)\;
  prior $\leftarrow$ ($\boldsymbol{\pi}^0,\boldsymbol{A}^0,\boldsymbol{\Theta}^0$)\;
  \For{$t\leftarrow {m+1}$ \KwTo $m+n$}{
    $\boldsymbol{P}^t \leftarrow concatenate(\boldsymbol{P}^{t-1},p_t)$\;
    \eIf{prior is None}{
        $\boldsymbol{\pi}^0,\boldsymbol{A}^0,\boldsymbol{\Theta}^0 \leftarrow $agile BNP-HMM$ (\boldsymbol{P}^t)$\;
    }{
        $\boldsymbol{\pi},\boldsymbol{A},\boldsymbol{\Theta} \leftarrow $agile BNP-HMM$(\boldsymbol{P}^t,prior$)\;
    }
    Delete the useless hidden states according to $\mathbb{E}[\boldsymbol{a}_j]<\gamma_u$\;
    Update the statistic $D_p$ according to \eqref{statistic}\;
    
    \eIf{$D_p \geq \gamma_p$}{
        $\boldsymbol{d}_{t} \leftarrow 1; prior \leftarrow None$\;
    }{
        $\boldsymbol{\pi}^0,\boldsymbol{A}^0,\boldsymbol{\Theta}^0 \leftarrow \boldsymbol{\pi},\boldsymbol{A},\boldsymbol{\Theta}; prior \leftarrow posterior$\;
    }
    }
    return $\boldsymbol{d}$\;
\end{algorithm}

\begin{table}
    \caption{Information on the Six Simulation Datasets }
	\centering
    \label{datasets}
	\begin{tabular}{cccl}
	    \hline
	    \makecell[c]{Simulation \\category} & Dataset & \makecell[c]{Modulation \\type} & \makecell[c]{Modulation\\ parameters ($\mu s$)}\\
	    \hline
		\multirow{5}*{\makecell[c]{PE\\task}} & \multirow{5}*{\makecell[c]{D1}} & 
		Agile & WM = [100,110,120,130,140]\\
		 &{~}& Staggered & WM = [100,110,120,130,140]\\
		 &{~}& Sliding & WM = (100, 140, 5)\\
		 &{~}& Jittered & WM = (125,5)\\
		 &{~}& D\&S & WM = (100, 140, 5)\\

		\hline
		\multirow{16}*{\makecell[c]{CPD\\ task\\(STVP)}}&
		\multirow{4}*{\makecell[c]{D2}}&
		\multirow{4}*{\makecell[c]{Staggered}}&
		WM1 = [100,110,115]\\
		&{~}&{~}&WM2 = [60,80,100,110]\\
		&{~}&{~}&WM3 = [70,75,88]\\
		&{~}&{~}&WM4 = [20,30,80]\\
		\cline{2-4}
		{~}&
		\multirow{4}*{\makecell[c]{D3}}&
		\multirow{4}*{\makecell[c]{Sliding}}&
		WM1 = (50, 110, 8)\\
		&{~}&{~}&WM2 = (50, 110, 6)\\
		&{~}&{~}&WM3 = (50, 80, 4)\\
		&{~}&{~}&WM4 = (50, 110, 3)\\
		\cline{2-4}
		{~}&
		\multirow{4}*{\makecell[c]{D4}}&
		\multirow{4}*{\makecell[c]{Agile}}&
		WM1 = [100,120,130]\\
		&{~}&{~}&WM2 = [60,80,100,110]\\
		&{~}&{~}&WM3 = [30,75,100]\\
		&{~}&{~}&WM4 = [50,60,70]\\
		\cline{2-4}
		{~}&
		\multirow{4}*{\makecell[c]{D5}}&
		\multirow{4}*{\makecell[c]{Jittered}}&
		WM1 = (100, 5)\\
		&{~}&{~}&WM2 = (150,5)\\
		&{~}&{~}&WM3 = (180,5)\\
		&{~}&{~}&WM4 = (200,5)\\
		\cline{1-4}
		\multirow{4}*{\makecell[c]{CPD\\task\\(VTVP)}}&
		\multirow{4}*{\makecell[c]{D6}}&
		Staggered &
		WM1 = [100,120,130,150,160]\\
		&{~}&Sliding&WM2 = (50, 150, 4)\\
		&{~}&Agile&WM3 = [100,120,130,150,160]\\
		&{~}&Jittered&WM4 = (125, 5)\\
		\hline
	\end{tabular}
\end{table}
\subsubsection{Evaluation Metrics}
The annotation error and Hamming distance are used to evaluate the \textbf{parameter estimation} performance.

\textbf{Annotation Error} $\Delta K$:
The annotation error reflects the ability of the algorithm to estimate the hidden states number: 
\begin{equation}
    \Delta K = \hat{K}-K
\end{equation}
wherein $\hat{K}$ and $K$ are the estimated and the true number of hidden states, respectively. ``$\Delta K$ close to zero" implied the algorithm achieved accurate estimations of hidden state numbers.

\textbf{Hamming Distance}: This metric indicates the ability to estimate an underlying state sequence, and the smaller its value is, the higher the estimation accuracy is. The Munkres algorithm~\cite{munkres1957algorithms} is used to map randomly selected indices of the estimated state sequence to the set of indices that maximized the overlap with the true sequence.

In the \textbf{change point detection} task, the performance is evaluated by following five metrics: F1 score, MDD, MT2FA, FAR, and MR.

\textbf{F1 Score:}
The F1 score is given in~\cite{romanenkova_indid_2022}, which reflects the detection accuracy.
\begin{equation}
    F1=\frac{TP}{TP+0.5(FP+FN)}
\end{equation}

\textbf{Mean Detection Delay (MDD) and Mean Time to False Alarm (MT2FA)}:
The MDD and MT2FA are introduced by Lordon\cite{lorden_procedures_1971}, and they are respectively expressed by \eqref{equ_5} and \eqref{equ_6}.

\textbf{False Alarm Rate (FAR) and Missing Rate (MR)}:
False alarm rate and missing rate are two typical metrics that represent the probability of incorrect detection~\cite{kay1993fundamentals}. 

\subsubsection{Baseline Methods}
Four change point detection methods are used as baseline methods in this study:
\begin{enumerate}
\item[1.] \textbf{Agile BNP-HMM-FSS} is implemented with the combination of the proposed parameter estimation method (agile BNP-HMM) and the sequential change point strategy of $\chi^2$-Fixed-Size-Sample~($\chi^2$-FSS)\cite{nikiforov_quadratic_1999}. The stopping time of $\chi^2$-FSS is calculated by:
\begin{equation}
    N = \inf \{mj:d_j=1\};\ 
    d_j=\begin{cases}
    1,\ $if$~|\chi_{(j-1)m+1}^{jm}|\geq h\\
    0,\ $if$~|\chi_{(j-1)m+1}^{jm}|< h
    \end{cases}
    \label{FSS}
\end{equation}
wherein $\chi_{(j-1)m+1}^{jm}$ is given by \eqref{chi_2}, $m$ is the fixed-size and $h$ is the threshold.
\item[2.] \textbf{U-FSS}\cite{bao_online_2022} has been used in our previous work on the MFR work mode online change point detection. It combines the generalized likelihood ratio test for the PE task and the original FSS algorithm\cite{nikiforov_two_1997} for the CPD task.
\item[3.] \textbf{U-CUSUM} is proposed in \cite{bao_online_2022} to fulfill the requirements of the MFR work mode online change point detection task. It combines the generalized likelihood ratio test for the PE task and the original CUSUM algorithm\cite{nikiforov_two_1997} for the CPD task.
\item[4.] \textbf{ChangeFinder} is a unifying framework for detecting outliers and change points from time series~\cite{takeuchi2006unifying}. It combines the auto-regressive model for the PE task and the logarithmic loss for the CPD task. It is used for the comparison under the non-ideal observations in this paper.
\end{enumerate}

For the simplicity of representation, in the following, ABHC is short for Agile BNP-HMM-CUSUM, ABHF is short for Agile BNP-HMM-FSS, and CF is short for ChangeFinder. 

\subsection{Functional Validation of Parameter Estimation}
\label{subsec:PEtask}
Two functionality of the PE task are evaluated in this section. Metrics are computed on a per-dataset basis and averaged over 100 Monte Carlo simulations. The hyper-parameters of the original BNP-HMM in the simulations are as follows: $\alpha_\pi=1,\alpha_A=1,a_0=1, b_0=0.01, \xi_0=0$, and $\lambda_0=1$. $\gamma_u$ is set as 0.1 for all datasets. To isolate the effects of the agile hyper-parameter setting, the same architecture and prior distributions are used for the subsequent model. In the results, some metrics are equal to zero and for presentation purpose these zeros results are assigned a minor positive value (0.005 in this study). Note that in the subsequent representation, the agile hyper-parameter are normalized so that $\kappa \in [0,1]$.
\subsubsection{Hidden State Number Estimation}
\begin{figure}[!t]
\centering
\includegraphics[width=3.2in]{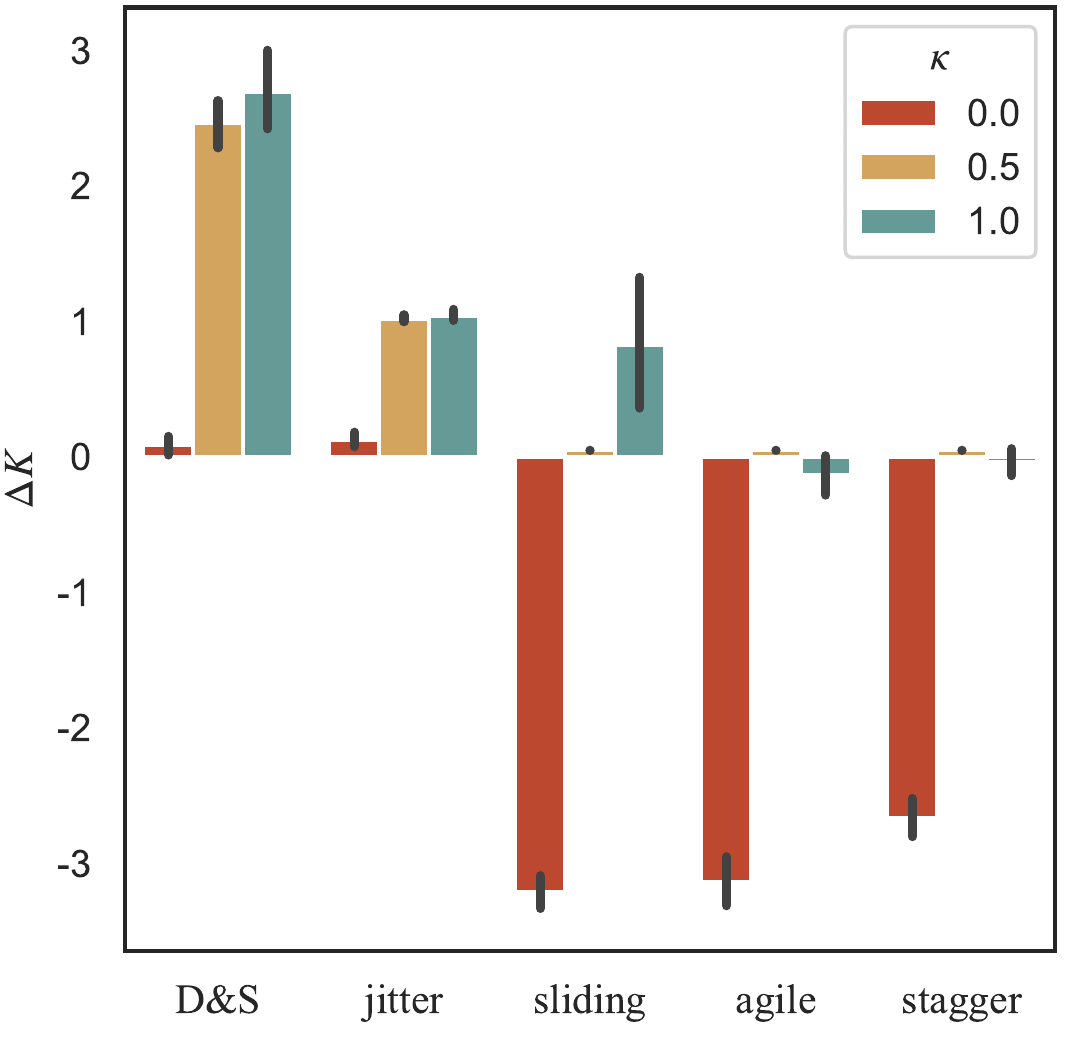}
\caption{$\Delta K$ of each modulation type estimated by the agile BNP-HMM with different $\kappa$ values.}
\label{deltak_five_types}
\end{figure}

The first simulation examines the influence of parameter $\kappa$ settings for both agile and non-agile inter-pulse modulation types. Agile BNP-HMM with $\kappa=0,0.5,1.0$ is used for five modulation types on D1. The estimated results of the hidden state number under different $\kappa$ settings are presented in Fig.~\ref{deltak_five_types}. In the simulations, the BNP-HMM with hyper-parameter $
\kappa=0.5, 1$ outperform that with $\kappa=0$ on the agile inter-pulse modulation types. Noted that this study focuses on the parameter estimation for agile inter-pulse modulation types. In subsequently reported simulations, the agile BNP-HMM with $\kappa=0.5$ is used on agile inter-pulse modulation types, and that with $\kappa=0$ was used for comparison.

\begin{figure}[!t]
\centering
\includegraphics[width=3.4in]{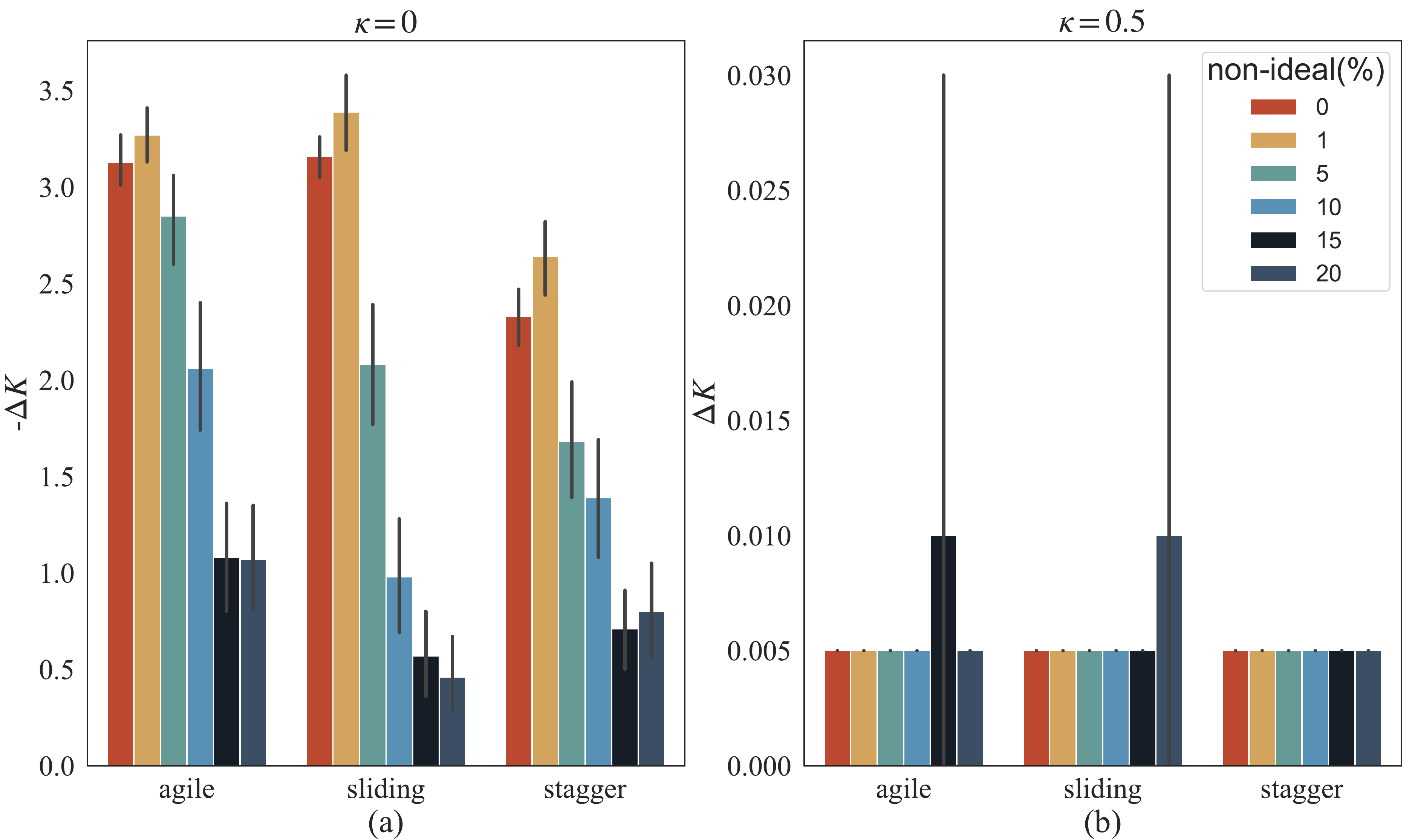}
\caption{$\Delta K$ performance versus the non-ideal ratio. Agile BNP-HMM results for D1 at: (a) $\kappa=0$ and (b) $\kappa=0.5$.}
\label{deltak_non_ideal}
\end{figure}

The performance under the non-ideal observations is presented in Fig.~\ref{deltak_non_ideal} (a) and~\ref{deltak_non_ideal} (b). The agile BNP-HMM performs better with $\kappa=0.5$ than $\kappa=0$. The $\Delta K$ is close to zero when $\kappa=0.5$ under various non-ideal settings. Note that in Fig.~\ref{deltak_non_ideal} (a), the agile BNP-HMM with $\kappa=0$ tends to underestimate the number of hidden states (the $y$-axis denotes $-\Delta K$), meanwhile the presence of more non-ideal pulses provides a more accurate hidden states assignments (the $\Delta K$ is close to zero when non-ideal ratio increase), which is the advantage of this method. 

\begin{figure*}[!t]
\centering
\includegraphics[width=7in]{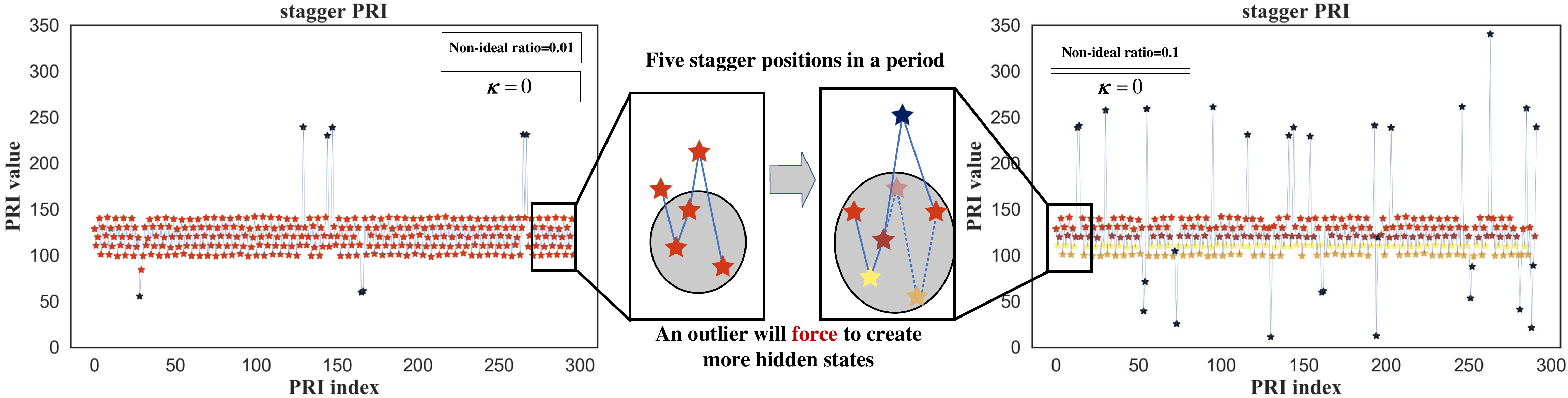}
\caption{The PE task results for different non-ideal ratios at $\kappa=0$.}
\label{various_in_non_ideal}
\end{figure*}

To analyze the reason of the phenomena, the PE task results of the non-ideal ratio values of 1\% and 10\% are shown in Fig.~\ref{various_in_non_ideal}. The hidden state labels inferred by the agile BNP-HMM are indicated by stars of different colors. The black stars represent $\Delta TOA$ values caused by missing and spurious pulses. The blue lines indicate the temporal relationship between different real PRI values and $\Delta TOA$ values. Theoretically, the original Dirichlet prior $(\kappa=0)$ encourages hidden states to have similar transition distributions $\mathbb{E}[a_{jk}|\alpha_A]$. A relatively high self-transition probability indicates that each PRI value belongs to a mutual hidden state, as shown in the left part of Fig.~\ref{various_in_non_ideal}. However, as shown in the right part of Fig.~\ref{various_in_non_ideal}, the difference between the $\Delta TOA$ values and real PRI values are large, thus the $\Delta TOA$ values could be easily identified as a new external hidden state with extremely low self-transition probability. Since outliers are randomly encountered and would violate the similar state transition assumption. When the detector encounters a $\Delta TOA$ value, the algorithm are more likely to move to another hidden state or create a new hidden state. By accumulating the transition tendency, more hidden states will be created. In the middle part, the detail of estimated hidden states from a pulse segment under 1\% and 10\% non-ideal ratio are shown. Only one hidden state is inferred under 1\% non-ideal ratio, and four hidden states are inferred under 10\% non-ideal ratio.


\subsubsection{Underlying State Sequence Estimation Performance}
\begin{figure}[!t]
\centering
\includegraphics[width=3.4in]{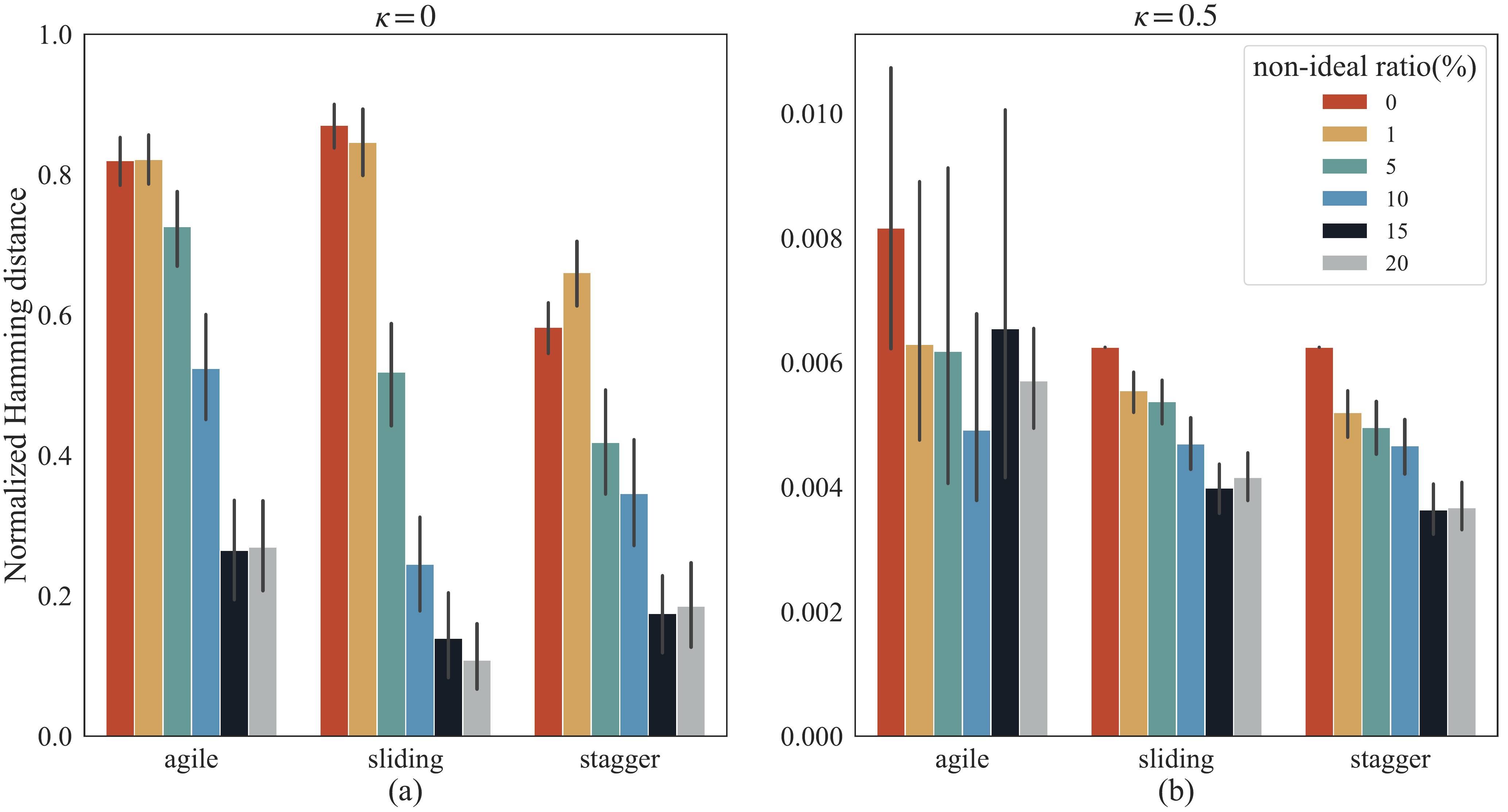}
\caption{Normalized Hamming distance with the non-ideal pulses in the scenario of agile modulation type. Agile BNP-HMM result for D1 at: (a)~$\kappa=0$ and (b)~$\kappa=0.5$}
\label{hamming_non_ideal}
\end{figure}

The second category of simulation is performed to evaluate the results of temporal relationship prediction. The normalized Hamming distance is calculated for the three typical agile inter-pulse modulation types at different non-ideal ratios. As shown in Fig.~\ref{hamming_non_ideal}~(a), with the increase in the non-ideal observation ratio, the agile BNP-HMM achieves better performance. The explanation is the same for the results in Fig.~\ref{various_in_non_ideal}.

From Fig.~\ref{hamming_non_ideal}~(a), the proposed model at $\kappa=0.5$ has a smaller normalized Hamming distance than the model at $\kappa=0$. There are two conclusions that can be drawn based on the results presented in Fig.~\ref{hamming_non_ideal} (b). Qualitatively, the agile BNP-HMM at $\kappa=0.5$ is robust to the non-ideal situation with the increase in the non-ideal ratio, and the normalized Hamming distance remains close to zero. The variance of agile PRI is larger than the other two modulation types. The hidden state transition distribution of agile PRI follows a uniform distribution, resulting in some miss-assignments. Quantitatively, the values of normalized Hamming distance  are between 0.005 and 0.01, implying that out of 1,000 pulses in the dataset. Five to ten pulses are assigned to wrong hidden states. The normalized Hamming distance is getting lower with the increase in the non-ideal ratio. 

Generally, we recommend setting the $\kappa$ to 0.5 when estimating the parameter of agile MFRs, and setting $\kappa$ to 0 when estimating the parameter of non-agile MFRs. In practical systems, we can not always obtain the ground truth to calculate $\Delta K$ and Hamming distance, thus we propose to perform the PE task on two branches with $\kappa$ values of 0.5 and 0. By calculating these branches in parallel, we can select the estimated value from the branch with the higher likelihood.

\subsection{Change Point Detection Performance Evaluation}
\label{subsec:CPDtask}

In change point detection tasks, datasets D2 to D6 are used for evaluation. In each simulation the initial-batch size was set to 20. The DPMM was initiated with the concentration parameter of $\alpha=1$ and the same concentration parameter as that used in~\cite{blei_variational_2006}.

\subsubsection{Basic Function Validation of Change Point Detection}
This section evaluates the performance of the proposed method and baselines under the optimal parameter settings. The parameters of the change detector are optimized based on the F1 score. As shown in Table~\ref{CPD_result}, the proposed ABHC method generally outperforms other baseline methods.

\begin{table}
    \caption{Quality Metrics of Different Change Point Detection Methods on Radar Data;
$\uparrow$ Indicates Metrics We want to Maximize; $\downarrow$ Indicates Metrics We want to Minimize; Best Values are Highlighted in Bold Font
\label{CPD_result}}
	\centering
	\begin{tabular}{lccccc}
	    \hline
	    Method & \makecell[c]{MDD \\(samples)$\downarrow$} & \makecell[c]{FAR \\(\%)$\downarrow$}  & \makecell[c]{MR\\ (\%)$\downarrow$}& F1 $\uparrow$ & \makecell[c]{MT2FA \\(samples)$\uparrow$} \\
	    \hline
	    \multicolumn{6}{c}{Only Staggered PRI/D2}\\
		\hline
		ABHC &\textbf{1.02}&	0.02&	\textbf{0.04}&	\textbf{0.96}&   $\boldsymbol{\infty}$ \\
		ABHF   &10.05&	\textbf{0.01}&	0.07&	\textbf{0.96}&	80.0 \\
		CF        &2.22&	0.45&	0.07&	0.67&	41.48\\
		U-FSS               &6.77	&0.303&	0.232&	0.722&	55.26\\
		U-CUSUM             & 5.43&	0.48&	0.39&	0.54&	127.42\\
		\hline
	    \multicolumn{6}{c}{Only Sliding PRI/D3}\\
	    \hline
		ABHC &\textbf{1.05}&	\textbf{0.05}&	\textbf{0.05}&	0.708&  130.5 \\
		ABHF   &15.92&	\textbf{0.05}&	0.27&	\textbf{0.86}&   $\boldsymbol{\infty}$\\
		CF        &3.35&	0.71&	0.53&	0.31&	114.83\\
		U-FSS               &6.34&  0.50&	0.31&	0.55&	21.19\\
		U-CUSUM             & 8.22&	0.4&    0.2&	0.67&	113.77\\
		\hline
	    \multicolumn{6}{c}{Only Agile PRI/D4}\\
	    \hline
		ABHC & \textbf{1.07}&	\textbf{0.09}&	\textbf{0.06}&	\textbf{0.94}&	\textbf{110.30} \\
		ABHF   &12.71&	0.14&	0.29&	0.79&	104.10\\
		CF        &1.23&	0.78&	0.46&	0.27&	84.47\\
		U-FSS               &6.6&	0.48&	0.40&	0.53&	94.83\\
		U-CUSUM             &8.43&	0.50&	0.17&	0.60&	68.79\\
		\hline
	    \multicolumn{6}{c}{Only Jittered PRI/D5}\\
	    \hline
		ABHC &0.75&	\textbf{0.00}&	\textbf{0.00}&	\textbf{1.00}&    $\boldsymbol{\infty}$\\
		ABHF   &11.7&  \textbf{0.00}&	0.08&	0.96&   $\boldsymbol{\infty}$\\
		CF        &\textbf{0.06}&	0.21&	\textbf{0.00}&	0.88&	53.31\\
		U-FSS               &9.25&	\textbf{0.00}&	\textbf{0.00}&   \textbf{1.00}&	$\boldsymbol{\infty}$\\
		U-CUSUM             &1.57&	\textbf{0.00}&	\textbf{0.00}&   \textbf{1.00}&	$\boldsymbol{\infty}$\\
		\hline
	    \multicolumn{6}{c}{Mixed modulation type/D6}\\
	    \hline
		ABHC &1.94&	\textbf{0.02}&	\textbf{0.02}&	\textbf{0.96}&	$\boldsymbol{\infty}$\\
		ABHF   &20.31&	0.04&	0.22&   0.88&	72.35\\
		CF        &\textbf{1.12}&	0.33&	0.54&	0.57&	11.00\\
		U-FSS               &4.62&	0.4&	0.23&	0.59&	58.66\\
		U-CUSUM             &7.17&	0.51&	0.53&	0.48&	128.6\\
		\hline
  
	\end{tabular}
\end{table}

The results validate the robustness of the proposed framework. As discussed in Section~\ref{subsec:PEtask}, the agile BNP-HMM with $\kappa=0.5$ achieves slightly inferior results in estimating the true number of hidden states and hamming distance for non-agile inter-pulse modulation types (such as Jittered PRI). In the change point detection, either the ABHC or the ABHF has an superior performance. Although the agile BNP-HMM may not always accurately identify all hidden states, it can successfully abstract the time series dynamics of different radar work modes, thus the performance of the CPD task is not degraded facing the non-agile work mode.

In practical applications, we have considered two scenarios. In the first scenario, there are some pre-accumulated pulse sequences with change point information analyzed and recorded by experts. Due to the existence of electronic intelligence these datasets are generally available. In this scenario, the ``optimal'' parameters can be optimized by some pre-defined objectives, such as the F1 score. These parameters are then used for subsequent pulse sequences. In the second scenario, there is no available dataset like in the first scenario. For instance, the MFR is an unknown MFR or a known MFR with unknown or software defined signals. In this case the corresponding radar signals may not be recorded by interceptor. In this scenario, parameters are directly defined and used for all change point detection tasks.


\subsubsection{Performance under Various Thresholds}
\begin{figure*}
    \centering
    \setkeys{Gin}{width=0.33\linewidth}
        \subfloat[D2]{\includegraphics{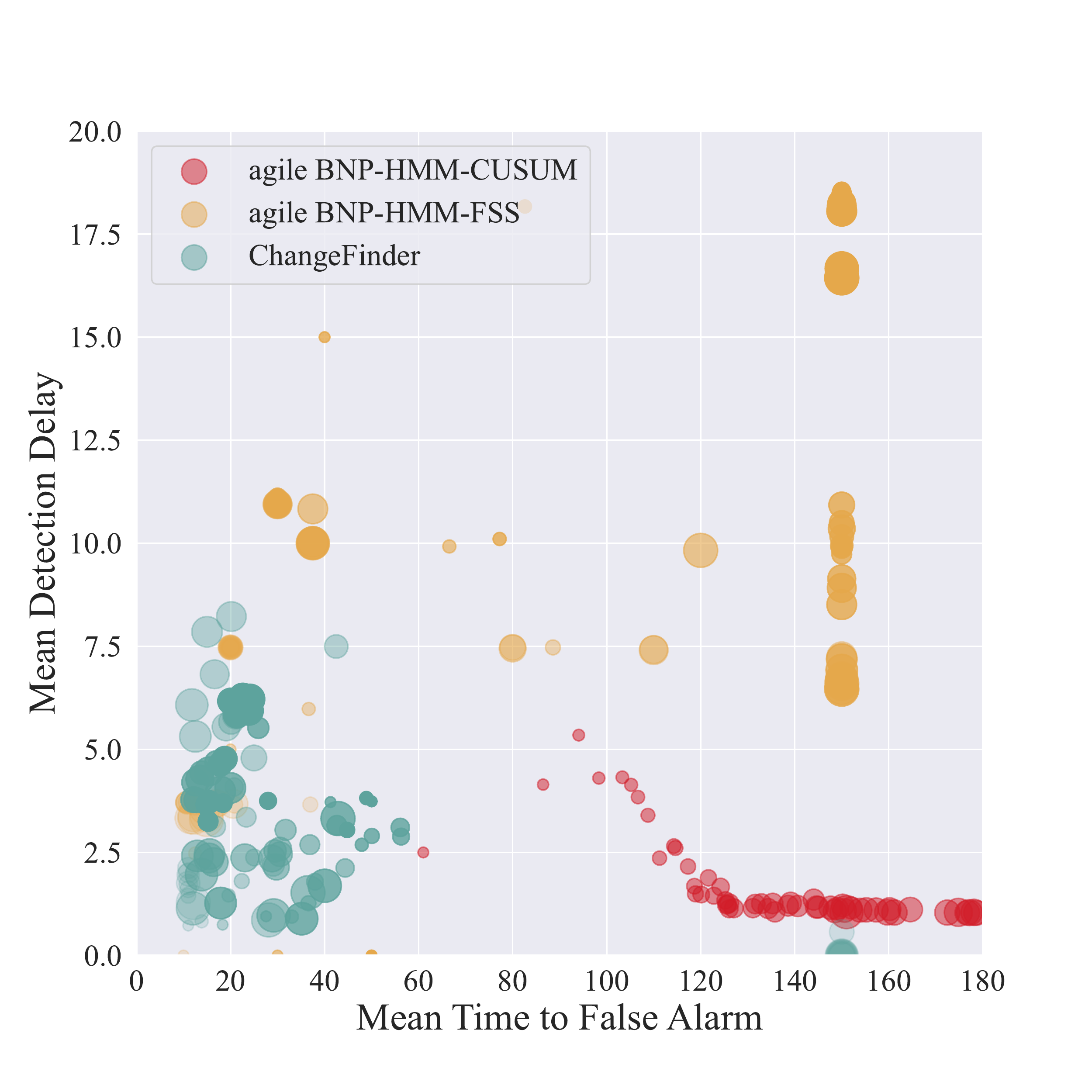}}
        \hfill
        \subfloat[D3]{\includegraphics{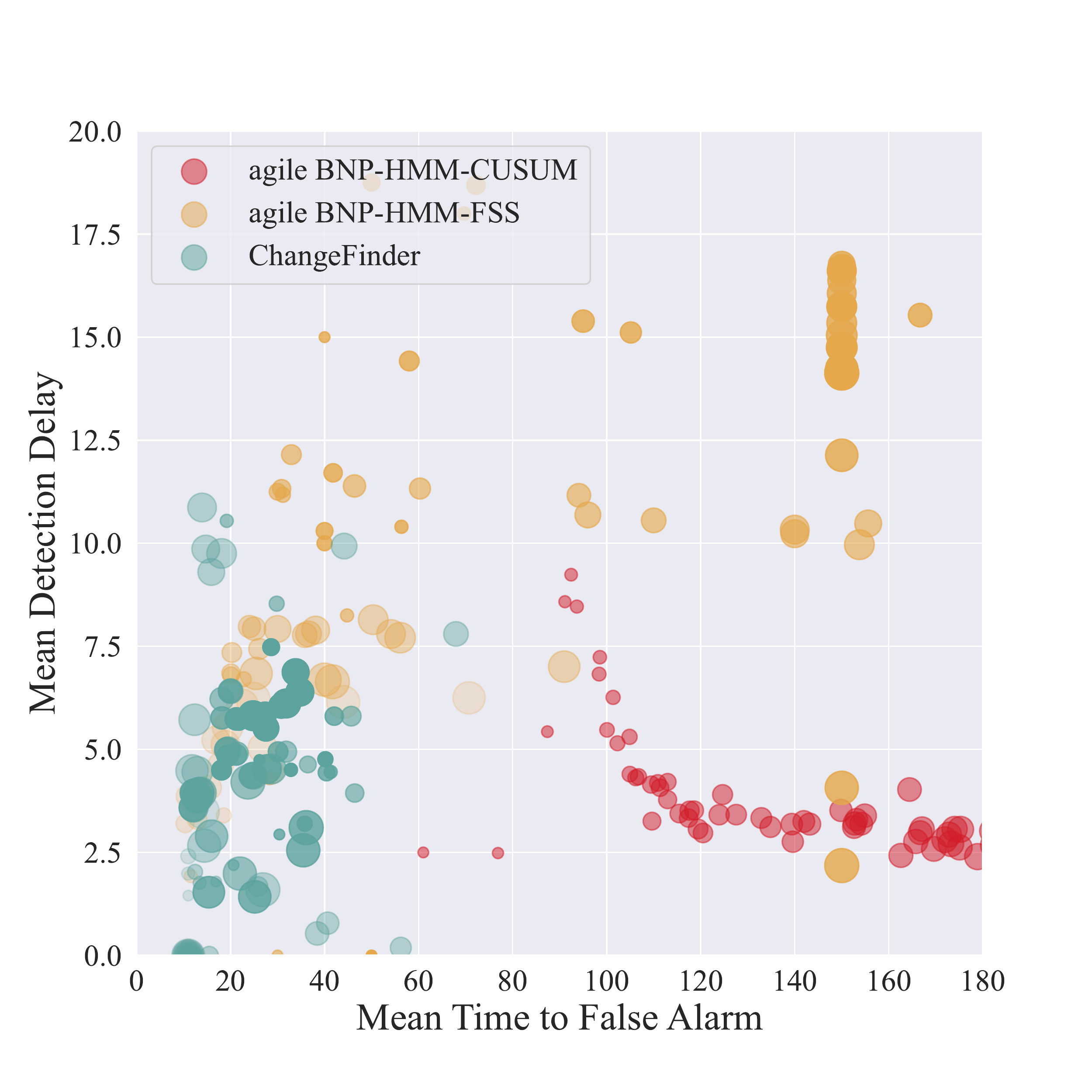}}
        \hfill
        \subfloat[D4]{\includegraphics{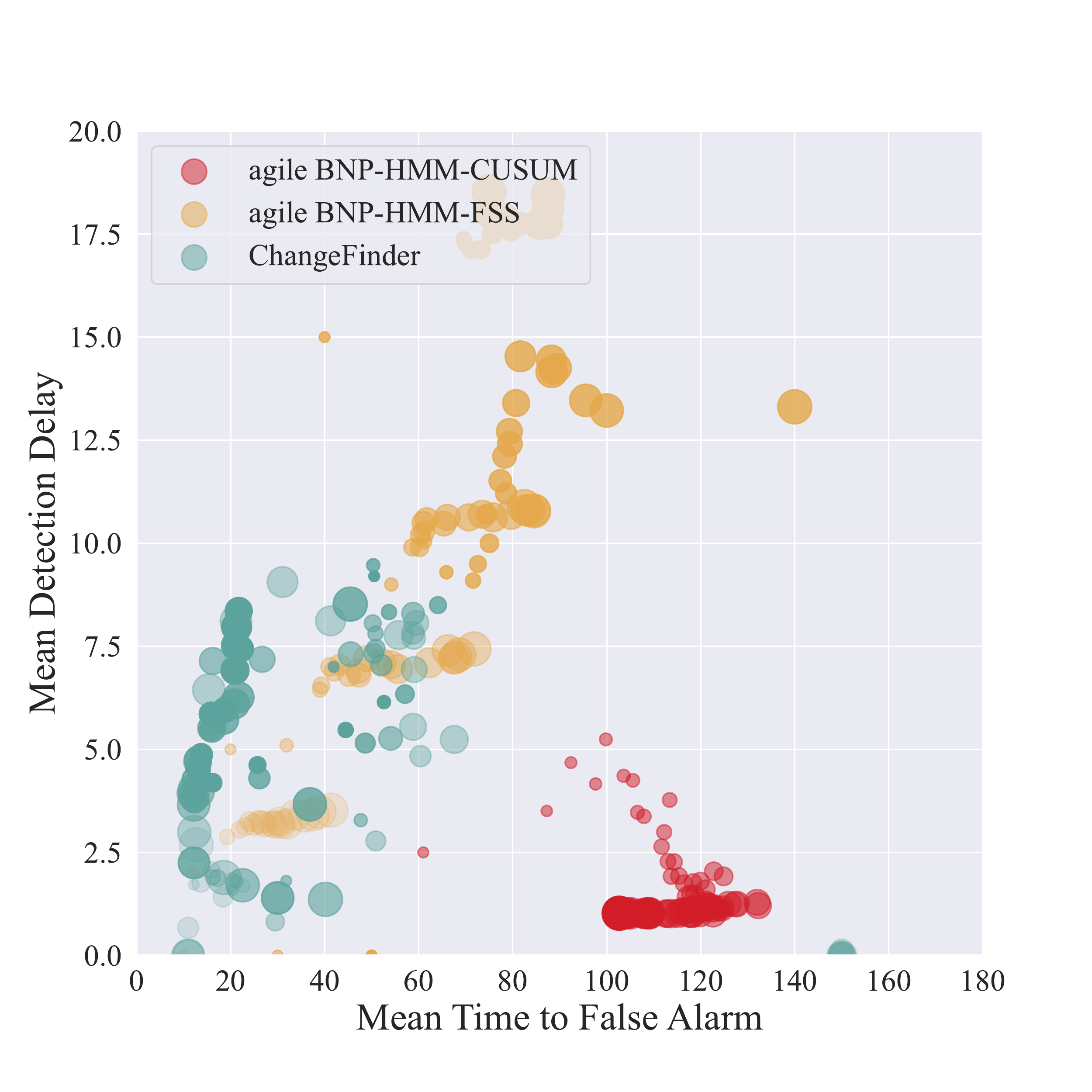}}
        
        \subfloat[D5]{\includegraphics{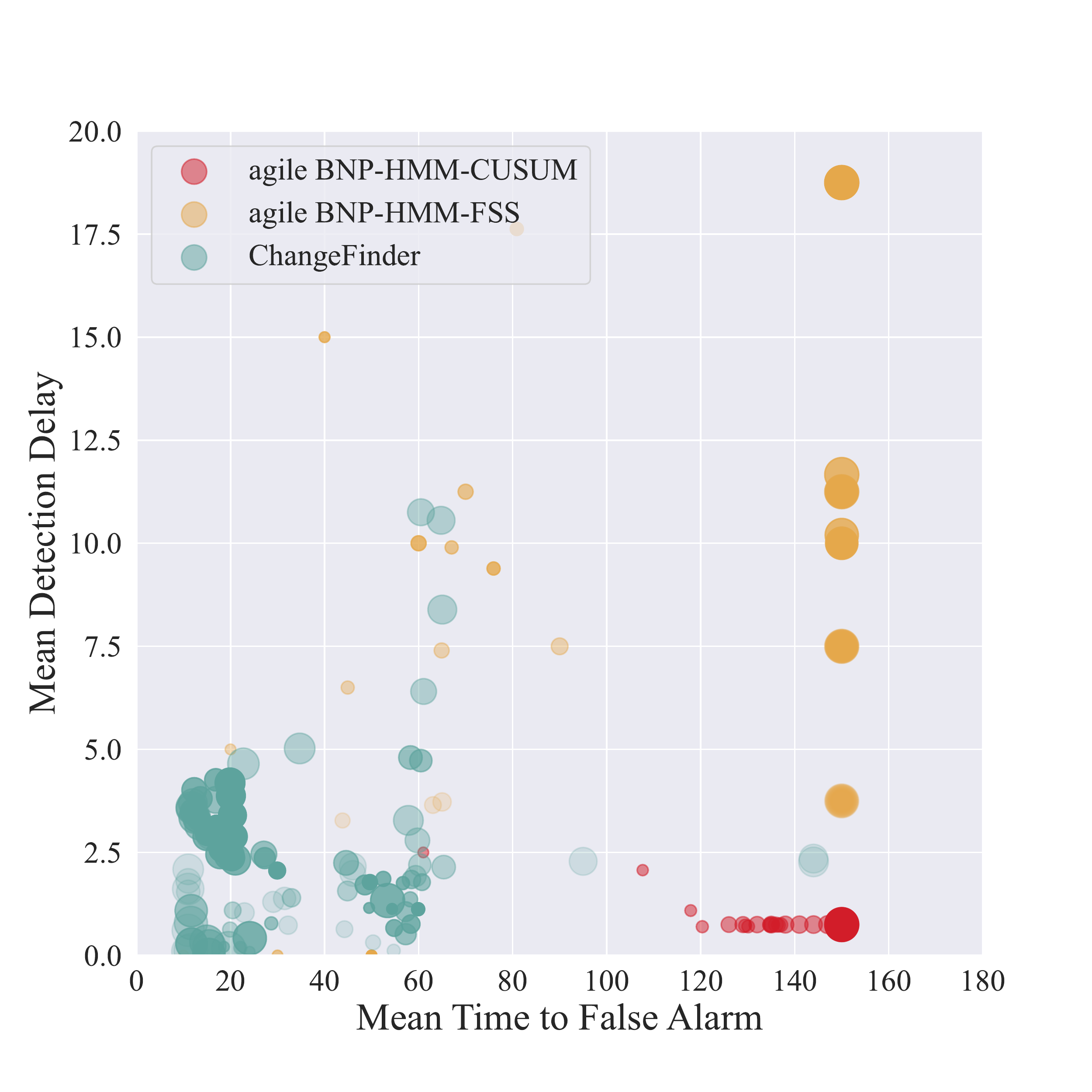}}
        \hfil
        \subfloat[D6]{\includegraphics{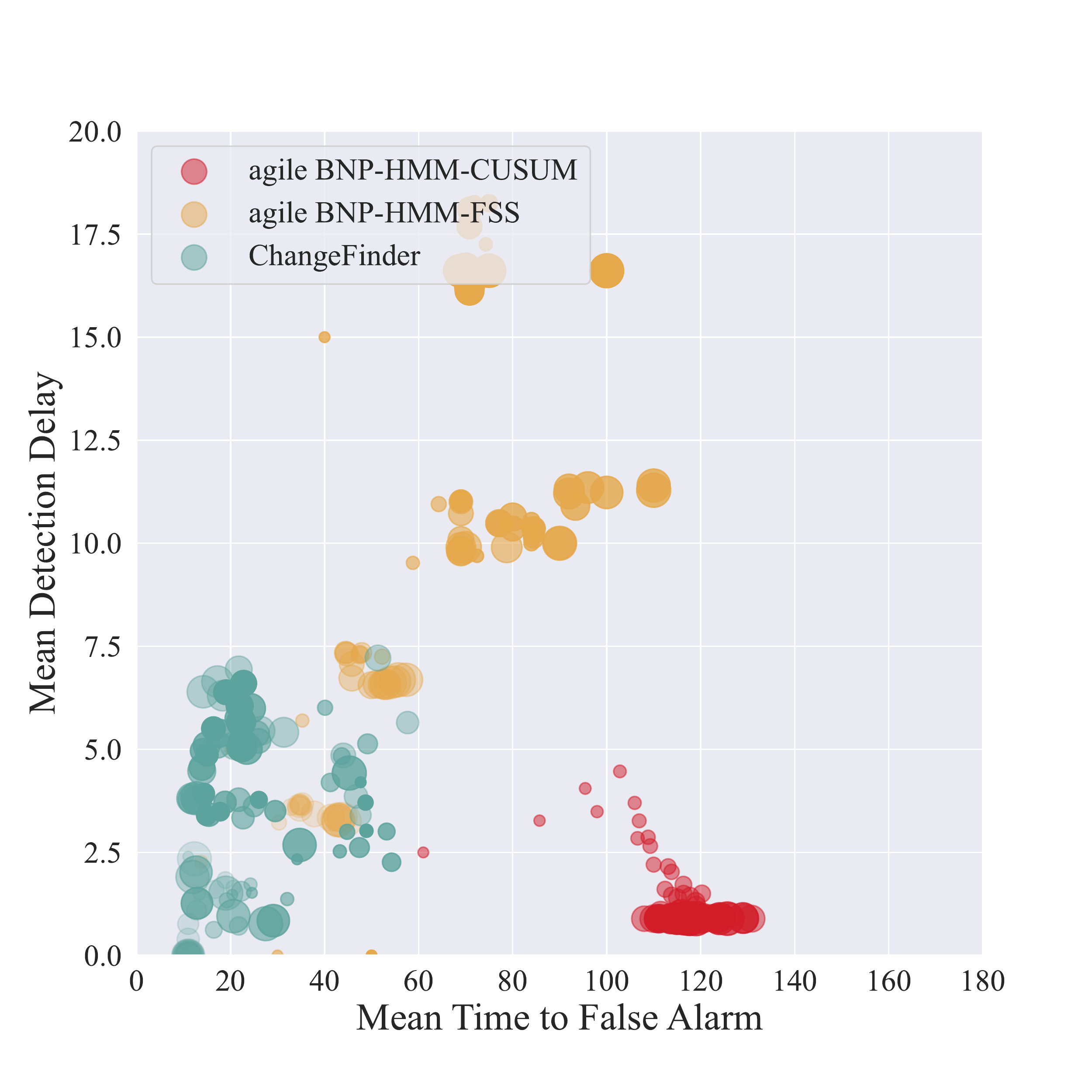}}
\caption{The MT2FA versus MDD detection scatter of the change point detection methods on different datasets: (a) D2; (b) D3; (c) D4; (d) D5; (e) D6. For the simplicity of representation, the case without any false alarm was assigned as MT2FA = 150. However, the maximum value of MT2FA was not 150.}
\label{detection_scatter}
    \end{figure*}

In this section, a detection scatter plot, as shown in Fig.~\ref{detection_scatter}, is designed to present the trade-off between MDD and MT2FA by tuning the change detector parameters. The $x$-axis represents the MT2FA, and the $y$-axis denotes the MDD. The feature of a scatter can be represented by a tuple $(\mathcal{T},\mathcal{S})$, wherein $\mathcal{T}$ and $\mathcal{S}$ represent the transparency and the size of a scatter, respectively. Since U-FSS and U-CUSUM is lack of interpretation, they are not simulated in the subsequent simulations. Fig.~\ref{detection_scatter} illustrates the influence of the thresholds selection on the quality metrics of the change point detection approaches. The parameters and their mapping on tuples were defined: 
\begin{equation}
\label{parameter mapping}
    (\mathcal{T},\mathcal{S})=
    \begin{cases}
    (r, k), & $if$ \    \rm $ChangeFinder$\\
    (m, h), & $if$ \    \rm $agile\ BNP-HMM-FSS$\\
    (None, \gamma_p),& $if$ \    \rm $agile\ BNP-HMM-CUSUM$
    \end{cases}
\end{equation}
wherein $(r,k)$ represents the decay and lagging factors of the CF; $(m,h)$ is the fixed-size and the threshold of the ABHF; and $\gamma_p$ represents the threshold of the ABHC. We first define the parameter range $(\mathcal{T}_{min},\mathcal{T}_{max}), (\mathcal{S}_{min},\mathcal{S}_{max})$. The candidate parameters are generated using equidistant sampling.

From the perspective of the scatter location, the major scatters of CF (green) are located at the bottom left of every sub-figure of Fig.~\ref{detection_scatter}, which implies a high false alarm rate. Meanwhile, most scatters of the ABHC framework (red) are located at the bottom right of the each sub-figure of Fig.~\ref{detection_scatter}, which demonstrates better timing performances. As for the ABHF framework, the specific design of the parameter estimation task ensures better performance at MT2FA. In some cases (see Fig.~\ref{detection_scatter} (a), \ref{detection_scatter} (b), and \ref{detection_scatter} (d)), the MT2FA value reaches 150, which indicated the zero false alarm rate on these data samples. The FSS strategy uses the data window for data points accumulation and decides whether there is a change point at the end of the data window (see~\eqref{FSS}), resulting in longer detection delays. The CUSUM strategy makes decisions for each incoming data point (see~\eqref{CUSUM}), resulting in smaller detection delay. 

From the perspective of parameter settings, the ABHC framework is more suitable for practical applications than the baseline methods. As shown in Fig.~\ref{detection_scatter} (e), the CF reconstructs the data via Auto-Regressive (AR) model, which is commonly used for modeling the stationary time series, causing the un-interpretable pattern for agile inter-pulse modulation. For ABHF, the scatters can be divided into five groups in terms of MDD as five $m$ were set in the simulations, reflecting the trade-off between the MDD and MT2FA. The transparency of each group from the bottom to the top decreased. The higher the fixed-size $m$ was set, the higher MT2FA and MDD were, achieving a trade-off between them. The size of each scatter increases in terms of the MT2FA value, indicating that a higher threshold $h$ would result in a higher MT2FA value. For ABHC framework, there is only one threshold parameter to be tuned, and the MT2FA increases (the false alarm rate decreased) with the scatter size increases (i.e., by setting a higher threshold). Varying the parameters would result in a stable MDD at a relatively low level, and the concentration of the ABHC scatters are much higher than in other methods.

\section{Experimental Performance on real-life Data}
\label{sec:experimental performance on real-life datasets}
The proposed method has the potential to other problems of detecting change points in highly structured time series, especially for the case wherein the pre- and post-change distribution can be well modeled by state space models. We evaluate the performance of our method on other two real-life signal processing benchmark datasets described below:\\
\textbf{HASC-2011}: It is a dataset of HASC challenge 2011 dataset \cite{kawaguchi2011hasc2011corpus}. It monitors human activity data via three-axis accelerometers. Six activities carried out are staying, walking, jogging, skipping, taking stairs up, and taking stairs down. The three-dimensional data was converted to one-dimensional data via $l^2$ norm, and then we quantify the sampled data so that the time series has discrete and finite values. Human activity recognition data is commonly used in CPD literature~\cite{giordano_evaluating_2018}.\\
\textbf{Well log}: It is a dataset of nuclear magnetic resonance measurements taken from a drill while drilling a well \cite{ruanaidh1994recursive}. It can be seen as a mean-shift Gaussian with outliers time series since the change in the rock stratification will produce the change in the mean of a time series. Change point detection algorithms applied to this dataset include \cite{giordano_evaluating_2018,knoblauch2018doubly}.

The experimental results of the real-life benchmark data can be found in Table \ref{result_real_data}. In HASC-2011, the ABHC method can capture the temporal feature of each motion and achieves the best FAR and MDR while maintaining the highest F1 score. The results of U-FSS, U-CUSUM, and CF methods are relatively poor because the change point detection method is not designed for time-varying series. The pre- and post-change distribution in well log dataset, can be modeled as a mean-shifted Gaussian time series with outliers. The ABHC method achieves better performance than the ABHF method on F1 score, as ABHC can learn the pattern from the beginning of the time series. The ABHF only learns from partial data (the data window). The U-FSS and U-CUSUM algorithm is sensitive to outliers and achieves worse results. The CF method also achieves better results as it can detect outliers and change point at the same time.

\begin{table}
    \caption{Quality metrics of different Change point detection methods on Real-life Data.
$\uparrow$ Indicates Metrics We want to Maximize; $\downarrow$ Indicates Metrics We want to Minimize; Best Values are Highlighted in Bold Font
\label{result_real_data}}
	\centering
		\begin{tabular}{lccccc}
	    \hline
	    Method & \makecell[c]{MDD \\(samples)$\downarrow$} & \makecell[c]{FAR \\(\%)$\downarrow$}  & \makecell[c]{MR\\ (\%)$\downarrow$}& F1 $\uparrow$ & \makecell[c]{MT2FA \\(samples)$\uparrow$} \\
		\hline
    \multicolumn{6}{c}{HASC-2011}\\
  \hline
		ABHC &88&	\textbf{0.38}&	\textbf{0.08}	&\textbf{0.70}&   $\boldsymbol{\infty}$ \\
		ABHF   &102&	0.61&	0.17	&0.53&	30\\
		CF        &45&	0.93&	0.23&	0.12&	25.64\\
		U-FSS               &51&	0.8&	0.83&	0.125&	85.26\\
		U-CUSUM            & 5.43&	0.93&	\textbf{0.01}	&0.47&	49.6\\
		\hline
	    \multicolumn{6}{c}{Well log}\\
	    \hline
		ABHC &100&	0.2&	\textbf{0.0}&	\textbf{0.94}&  $\boldsymbol{\infty}$ \\
		ABHF   &130&	0.3&	0.17	&0.6&   $\boldsymbol{\infty}$\\
		CF        &\textbf{64	}&\textbf{0.0}&	0.08	&0.9&	114.83\\
		U-FSS               &126&	0.9&	0.25	&0.55&	126\\
		U-CUSUM             & 176&	\textbf{0.0}&	0.1	&0.85&	150\\
        \hline
	\end{tabular}
\end{table}

\section{Conclusion}
\label{sec:conclusion}
 In this paper, a novel Bayesian non-parametric parameter-based framework for parameter estimation and change point detection of MFR work modes is proposed. Firstly, a fully conjugate generative model is designed, which enables highly efficient variational inference. Secondly, the Dirichlet process with an agile feature is designed, and a stick-breaking construction is proposed to control the tendency of Markov Chain self-transitioning. Thirdly, the variational iteration function and the error probability lower bound of the PE task is derived. The proposed parameter estimation method is further combined with streaming Bayesian updating to reduce the computational redundancy. Finally, the optimal sequential strategy $\chi^2$-CUSUM with agile BNP-HMM is designed for better change point detection performance.

The resulting framework do not require prior information, free of window setting dilemmas, easy to set parameters, and robust to non-ideal observations. Results showed that the proposed framework is highly competitive compared to other four methods on either simulated PRI datasets or real-life benchmark datasets.

There are some future works: 1) First-order Markov property is assumed in this study, which may limit the model representation ability; 2) The changes in either modulation type or parameters for modern MFR and cognitive radar is driven by higher level mission objectives. The results of our proposed method can support the inverse construction of the MFR and cognitive radars’ objective functions in the future.

\section*{Acknowledgments}
The authors appreciate the editors and anonymous referees for their efforts and constructive comments to improve the quality of this paper.



 

\bibliographystyle{IEEEtran}
\bibliography{ref}

\vfill

\end{document}